%%%%%%%%%%%%%%%%%%%%%%%%%%%%%%%%%%%%%%%%%%%%%%%%%%%%%%%%%%%%%%%%%%%%%%%%%%
%                Exact Meander Asymptotics: a Numerical Check 		 %
%									 % 
%                by P. Di Francesco, E. Guitter and J. Jacobsen          %
%                TEX file, using harvmac.tex macros  		         %
%									 %
%                    SPhT 00/027  					 %
%%%%%%%%%%%%%%%%%%%%%%%%%%%%%%%%%%%%%%%%%%%%%%%%%%%%%%%%%%%%%%%%%%%%%%%%%%
\input harvmac 
\input epsf.tex

\overfullrule=0mm

\newcount\figno \figno=0
\newcount\figtotno
\figtotno=0
\newdimen\captionindent
\captionindent=1cm

\newcount\figno
\figno=0
\def\fig#1#2#3{
\par\begingroup\parindent=0pt\leftskip=1cm\rightskip=1cm\parindent=0pt
\baselineskip=11pt
\global\advance\figno by 1
\midinsert
\epsfxsize=#3
\centerline{\epsfbox{#2}}
\vskip 15pt
{\bf Fig. \the\figno:} #1\par
\endinsert\endgroup\par
}
\def\figlabel#1{\xdef#1{\the\figno}}
\def\encadremath#1{\vbox{\hrule\h box{\vrule\kern8pt\vbox{\kern8pt
\hbox{$\displaystyle #1$}\kern8pt}
\kern8pt\vrule}\hrule}}
\def\encadre#1{\vbox{\hrule\hbox{\vrule\kern8pt\vbox{\kern8pt#1\kern8pt}
\kern8pt\vrule}\hrule}}

\def\em{\it}
\def\frac#1#2{{{#1}\over{#2}}}

%Macros 
%%%%%%%%%%%%%%%%%%%%%%%%%%%%%%%%%%%%%%%%%%%%%%%%%%%%%%%%%%%%%%%%%
\def\tvi{\vrule height 10pt depth 6pt width 0pt}
\def\tv{\tvi\vrule}

\def\mod{{\rm mod \ }}
\def\IR{\relax{\rm I\kern-.18em R}}
\font\cmss=cmss10 \font\cmsss=cmss10 at 7pt
\def\IZ{\relax\ifmmode\mathchoice
{\hbox{\cmss Z\kern-.4em Z}}{\hbox{\cmss Z\kern-.4em Z}}
{\lower.9pt\hbox{\cmsss Z\kern-.4em Z}}
{\lower1.2pt\hbox{\cmsss Z\kern-.4em Z}}\else{\cmss Z\kern-.4em Z}\fi}
\def\buildrel#1\under#2{\mathrel{\mathop{\kern0pt #2}\limits_{#1}}}
%%%%%%%%%%%%%%%%%%%%%%%%%%%%%%%%%%%%%%%%%%%%%%%%%%%%%%%%%%%%%%%%%

\Title{SPhT/00-027}
{{\vbox {
%\centerline{}
\medskip
\centerline{Exact Meander Asymptotics: a Numerical Check}
}}}
\medskip
\centerline{P.~Di Francesco\footnote*{e-mail:philippe@spht.saclay.cea.fr},}
\medskip
\centerline{E.~Guitter\footnote{$^\#$}{e-mail:guitter@spht.saclay.cea.fr},}
\medskip

\centerline{ \it Service de Physique Th\'eorique, C.E.A. Saclay,}
\centerline{ \it F-91191 Gif sur Yvette, France}

\medskip
\centerline{and} 
\medskip
\centerline{J.~L.~Jacobsen\footnote{$^\&$}{e-mail:jacobsen@ipno.in2p3.fr
},}
\medskip

\centerline{ \it LPTMS, b\^atiment 100, Universit\'e Paris-Sud,}
\centerline{ \it F-91405 Orsay, France}

\bigskip

{\bf Abstract:}
This note addresses the meander enumeration problem: ``Count all topologically
inequivalent configurations of a closed planar non self-intersecting curve 
crossing a line through a given number of points". We review a description of 
meanders introduced recently in terms of the coupling to gravity of a two-flavored 
fully-packed loop model. The subsequent analytic predictions for various meandric 
configuration exponents are checked against exact enumeration, using a transfer 
matrix method, with an excellent agreement.
\bigskip

\noindent
\Date{03/00}
%\draft
%\Date{Keywords: meanders, polymers, folding, matrix models.}
%\writetoc
%references

\nref\ASL{A.~Sainte-Lagu\"e, {\it Avec des nombres et des lignes
(R\'ecr\'eations math\'ematiques)}, Vuibert, Paris (1937).}

\nref\HMRT{K.~Hoffman, K.~Mehlhorn, P.~Rosenstiehl and R.~Tarjan, {\it
Sorting Jordan sequences in linear time using level-linked search
trees}, Information and Control {\bf 68} (1986) 170--184.}

\nref\ARNO{V.~Arnold, {\it The branched covering of $CP_2 \to S_4$,
hyperbolicity and projective topology},
Siberian Math.~Jour.~{\bf 29} (1988) 717--726.}

\nref\KOSMO{K.~H.~Ko and L.~Smolinsky, {\it A combinatorial matrix in
$3$-manifold theory}, Pacific.~J.~Math {\bf 149} (1991) 319--336.}

\nref\DGG{P.~Di Francesco, O.~Golinelli and E.~Guitter, {\it Meander,
folding and arch statistics}, Mathl.~Comput.~Modelling {\bf 26} (1997) 97--147.}

\nref\TOU{J.~Touchard, {\it Contributions \`a l'\'etude du probl\`eme
des timbres poste}, Canad.~J.~Math.~{\bf 2} (1950) 385--398.}

\nref\LUN{W.~Lunnon, {\it A map--folding problem},
Math.~of Computation {\bf 22}
(1968) 193--199.}

\nref\NOUS{P.~Di~Francesco, O.~Golinelli and E.~Guitter, {\it Meanders: 
a direct enumeration approach}, Nucl.~Phys.~B {\bf 482} [FS] (1996) 497--535.}

\nref\Goli{O.~Golinelli, {\it A Monte-Carlo study of meanders}, preprint 
cond-mat/9906329, to appear in Eur.~Phys.~J.~B (2000).}

\nref\JEN{I.~Jensen, {\it Enumerations of plane meanders}, preprint 
cond-mat/9910313 (1999).}

\nref\BIPZ{E.~Br\'ezin, C.~Itzykson, G.~Parisi and J.-B.~Zuber,
{\it Planar diagrams}, Comm.~Math.~Phys.~{\bf 59} (1978) 35--51.}

\nref\LZ{S.~Lando and A.~Zvonkin, {\it Plane and projective meanders},
Theor.~Comp.~Science {\bf 117} (1993) 227--241; {\it Meanders},
Selecta Math.~Sov.~{\bf 11} (1992) 117--144.}

\nref\MAK{Y.~Makeenko {\it Strings, matrix models, and meanders}, 
Nucl.~Phys.~Proc. Suppl.~{\bf 49} (1996) 226--237.}

\nref\SSz{G.~Semenoff and R.~Szabo {\it Fermionic matrix models},
Int.~J.~Mod.~Phys.~A {\bf 12} (1997) 2135--2292.}

\nref\CK{L.~Chekhov and C.~Kristjansen,
{\it Hermitian matrix model with plaquette 
interaction}, Nucl.~Phys.~B {\bf 479} (1996) 683--696.}

\nref\EKR{B.~Eynard and C.~Kristjansen, {\it More on the exact solution of the O(n)
model on a random lattice and an investigation of the case $|n|>2$},
Nucl.~Phys.~B {\bf 466} [FS] (1996) 463--487.}

\nref\GGD{P.~Di~Francesco, O.~Golinelli and E.~Guitter, {\it Meanders: Exact
asymptotics}, preprint cond-mat/9910453, to appear in Nucl.~Phys.~B (2000).}

\nref\JACO{J.~L.~Jacobsen and J.~Kondev, {\it Field theory of compact polymers
on the square lattice}, Nucl.~Phys.~B {\bf 532} [FS], (1998) 635--688;
{\it Transition from the compact to the dense phase of two-dimensional polymers}, 
J.~Stat.~Phys.~{\bf 96}, (1999) 21--48.}

\nref\KPZ{V.~G.~Knizhnik, A.~M.~Polyakov and A.~B.~Zamolodchikov,
{\it Fractal structure of 2D quantum gravity},
Mod.~Phys.~Lett.~A {\bf 3} (1988) 819--826;
F.~David, {\it Conformal field theories coupled to 2D gravity in the
conformal gauge}, Mod.~Phys.~Lett.~A {\bf 3} (1988) 1651--1656;
J.~Distler and H.~Kawai,
{\it Conformal field theory and 2D quantum gravity},
Nucl.~Phys.~B {\bf 321} (1989) 509.}

\nref\EKN{E.~Guitter, C.~Kristjansen and J.~Nielsen, 
{\it Hamiltonian cycles on random Eulerian triangulations},
Nucl.~Phys.~B {\bf 546} [FS] (1999) 731--750; 
P.~Di~Francesco, E.~Guitter and C.~Kristjansen,
{\it Fully packed O(n=1) model on random Eulerian triangulations},
Nucl.~Phys.~B {\bf 549} [FS] (1999) 657--667.}

\nref\progress{P.~Di Francesco, E.~Guitter and J.~L.~Jacobsen,
work in progress.}

\nref\KE{B.~Eynard and C.~Kristjansen, {\it 
An iterative solution of the three-colour problem on a random lattice},
Nucl.~Phys.~B {\bf 516} (1998) 529--542.}

\nref\DaDu{F.~David and B.~Duplantier, {\it Exact partition functions
and correlation
functions of multiple Hamiltonian walks on the Manhattan lattice},
J.~Stat.~Phys.~{\bf 51}, (1988) 327--434. H.~Saleur, J.~Phys.~A {\bf 19}
(1986) L807--L810.}

\nref\Nienh{B.~Nienhuis, in {\it Phase transitions and critical phenomena},
edited by C.~Domb and J.~L.~Lebowitz (Academic, London, 1987), Vol.~11.}

\nref\DSZ{P.~Di Francesco, H.~Saleur and J.-B.~Zuber, {\it Modular
invariance in non-minimal two-dimensional conformal theories},
Nucl.~Phys.~B {\bf 285} (1987) 454--480.}

\nref\SAL{H.~Saleur, {\it Polymers and percolation in two dimensions
and twisted N=2 supersymmetry}, Nucl.~Phys.~B {\bf 382} (1992) 486--531.}

\nref\Enting{I.~G.~Enting, {\it Generating functions for enumerating
self-avoiding rings on the square lattice},
J.~Phys.~A {\bf 13} (1980) 3713--3722.}

\nref\Derrida{B.~Derrida, {\it Phenomenological renormalisation of the
self avoiding walk in two dimensions}, J.~Phys.~A {\bf 14} (1981) L5--L9.}

\nref\Blote{H.~W.~J.~Bl\"{o}te and M.~P.~Nightingale,
{\it Critical behaviour of the two-dimensional Potts model with a
continuous number of states: A finite size scaling analysis},
Physica A {\bf 112} (1982) 405--465.}

\nref\LOR{P.~Di~Francesco, E.~Guitter and C.~Kristjansen, {\it Integrable
2D Lorentzian gravity and random walks}, preprint hep-th/9907084, to appear
in Nucl.~Phys.~B (2000).}

\nref\BAR{B.~Durhuus, J.~Fr\"olich and T.~J\'onsson,
Nucl.~Phys.~B {\bf 240} (1984) 453--480;
F.~David, Nucl.~Phys.~B {\bf 487} [FS] (1997) 633--649.}
 
\nref\LOW{G.~Lowler, O.~Schramm and W.~Werner, {\it Values of
Brownian intersection exponents I: half plane exponents}, preprint
math.PR/9911084 (1999).}

\nref\FLM{D.S.~Fisher, P.~Le Doussal and C.~ Monthus, 
{\it Random walkers in 1-D random environments: Exact renormalization group
analysis},  Phys.~Rev.~E {\bf 59} (1999) 4795--4840.}

%text 

\newsec{Introduction}

The meander problem is one of those tantalizing questions that has resisted
a definite solution for decades, although it is very easy to state:
``Given a positive integer $n$, in how many topologically distinct ways can
a closed non-intersecting planar curve ({\em road}) cross a straight
line ({\em river}) in exactly $2n$ points ({\em bridges})?''

Originally an exercise of recreational mathematics~\ASL, the meander problem
turned out to have applications in the most various branches of science:
Sorting algorithms in computer science~\HMRT, enumeration of ovals of planar
algebraic curves~\ARNO, classification of three-manifolds~\KOSMO, and in
connection with a particular type of self-avoiding walk describing the
compact folding of a linear chain~\DGG.

An obvious strategy would of course be to evaluate the first few meander
numbers $M_n$, in the hope of finding an explicit formula, valid for arbitrary $n$.
Such enumerative approaches exist on various levels of
sophistication~[\xref\TOU-\xref\Goli], and a recent transfer matrix method~\JEN\
carried out this program up to $n=24$. Although an explicit
expression for $M_n$ appears to be out of reach, it became clear that, in
analogy with two-dimensional lattice polymers, the meander numbers
scale asymptotically as $M_n \sim C R^{2n} / n^{\alpha}$, where $R$ is a
connectivity constant and $\alpha$ a configuration exponent.

A major achievement of random matrix theory has been to deal with precisely
such asymptotic enumeration problems~\BIPZ. It is therefore natural to
apply such techniques to the meander problem~[\xref\LZ-\xref\SSz]. In particular
it has emerged that a generalized multi-road multi-river meander problem,
in which each closed segment of river (resp.~road) is given the 
statistical weight
$n_1$ (resp.~$n_2$), can be cast as a Hermitian matrix model, known as the
O($n_1$,$n_2$) model~\DGG. In the special case of $n_2=1$ this model is
soluble by a saddle-point method, leading to an exact evaluation of
$R$ and $\alpha$~[\xref\CK,\xref\EKR]. Unfortunately these results do not
pertain to the original meander problem, which is recovered in the limit
$n_1,n_2 \to 0$.

In a recent publication  \GGD\ it was argued that the meander problem
is a particular realization of the coupling to gravity of
a certain two-flavored loop model~\JACO, initially defined on the square
lattice.
The most general gravitational version of this loop model is a 
generalization of the meander problem in which river and road
segments, counted with their respective weights of $n_1$ and $n_2$, are
allowed to cross as well as to touch one another without crossing
(tangency points) \GGD .
We shall refer to this model as {\em tangent meanders}. On the regular
lattice, directed segments of river and road can be inserted by means
of certain magnetic defect operators, for which the anomalous dimensions
are known exactly. When dressed by quantum gravity, these dimensions
transform according to the KPZ formula~\KPZ. This transformation allows
one to extract {\em exact} values for the configuration exponent $\alpha$ of
tangent meanders, whereas $R$, being a
non-universal quantity, is lost in the process.
The connection to the original meander problem is then made
by arguing that tangency is {\em irrelevant} from a
renormalization point of view \GGD. 
Thus, the result for $\alpha$ in fact pertains to the
original meander problem, i.e. to the gravitational O($n_1$,$n_2$) model.
Moreover, the operator content of the theory gives access to
other geometries, involving several rivers, possibly with marked points,
as well as semi-meanders (river with a source).

Here we review and extend the arguments of \GGD. In particular we
establish the irrelevance of tangency rigorously in a
number of special cases. 
We also add credibility to the theoretical predictions by performing
extensive exact enumerations of various meander geometries, using a
generalization of the transfer matrix method presented in~\JEN.
We first address multi-component meanders, that allow for precisely
checking the predicted value of the central charge of the underlying
conformal theory. Next we explore two distinct river geometries, namely
(i) two parallel rivers, and (ii) one semi-infinite river, that
permit to validate their magnetic operator formulation within
the corresponding conformal theory. We finally consider the case
of tangent meanders and verify the irrelevance of tangency,
thus confirming the cornerstone of the argument.
In all cases we find an excellent agreement with theory, typically confirming
the configuration exponents with 4--5 significant digits.

The paper is organized as follows.
In Section 2 we review the square-lattice loop model and its solution,
before presenting its gravitationally dressed version and the results
for the asymptotics of a range of meander-related quantities. The
transfer matrix algorithms are presented in Section 3, and in Section 4
we analyze the data and compare them to the theoretical predictions.
Our conclusions and some perspectives can be found in Section 5.

\newsec{Theory: from fully-packed loop gases to meanders}

In this Section, we review the arguments of Ref.~\GGD\ relating the
meander problem to the gravitational version of a particular
fully-packed loop model initially defined on the square lattice~\JACO.
The effect of gravity is to replace the lattice with a random 
quadrangulation of the sphere. The lattice loop gas is described
in subsection 2.1 while its conformal structure is presented in
subsection 2.2. The connection to meanders via two-dimensional quantum
gravity is explained in subsection 2.3. This leads naturally to
an effective field theory description detailed in subsection 2.4
together with the subsequent predictions for various meandric configuration
exponents.

\subsec{Fully-packed loop gas on the square lattice}

\fig{A typical fully-packed loop configuration on the square lattice. Assuming
doubly periodic boundary conditions, there are 6 black loops 
(solid lines) and 4 white ones (dashed lines). Up to rotations, the vertices of the
model are of the two types (a) ``crossing" or (b) ``avoiding".}{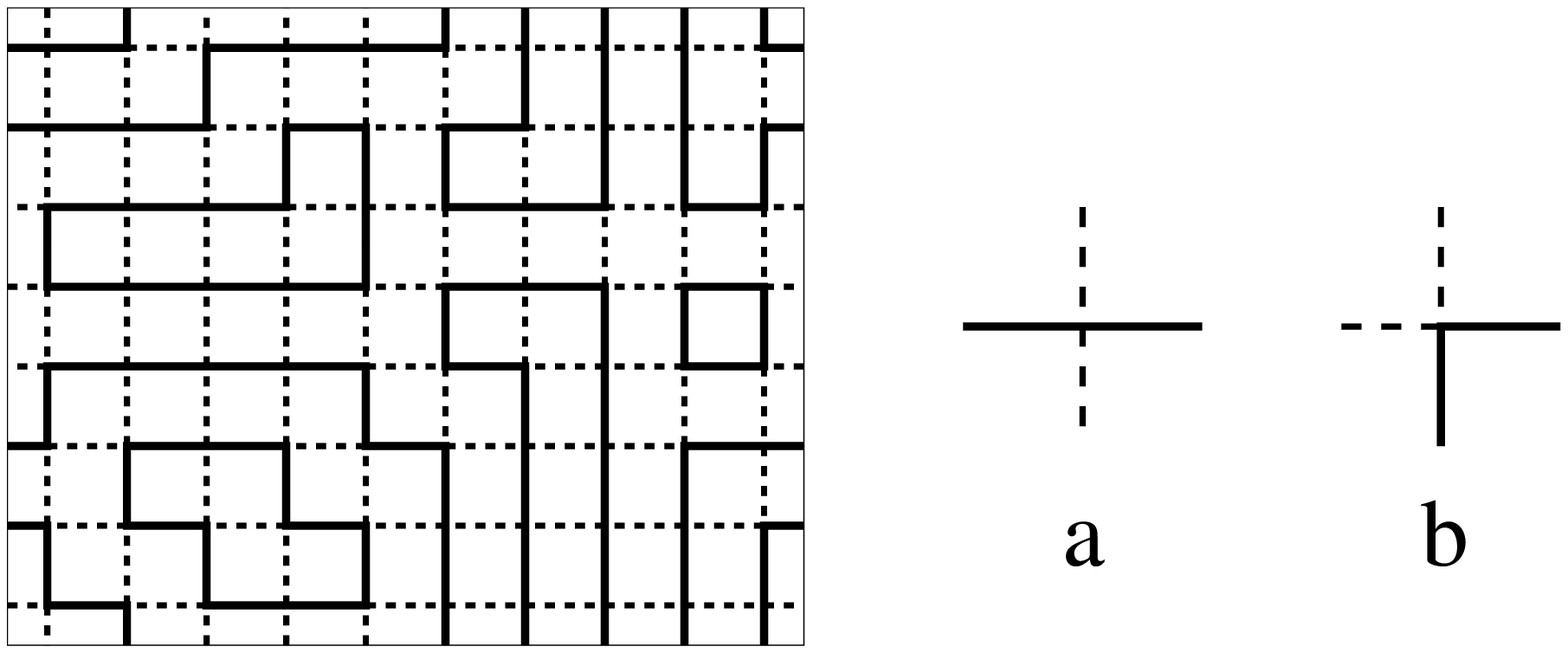}{8.cm}
\figlabel\twovert

The configurations of the fully-packed loop model that we shall consider
are defined by assigning to each
edge of the two-dimensional square lattice
either of two colors (say, black or white, represented as solid or dashed lines
in Fig.~\twovert, 
also referred to as $1$ and $2$ in the following)
in such a way that each vertex has exactly two black and two white incident edges.  
Up to obvious rotations, this gives rise to the two vertex configurations
depicted in Fig.~\twovert\ 
in which the black and white lines either avoid or cross each other.
Note that with periodic boundary conditions the black and white lines form loops.
It is interesting to remark that the fully-packed loop model's configurations 
defined here differ from 
those of the so-called densely packed 
loop model~\JACO\ in that each vertex is visited by a black {\it and} a white loop, whereas in the 
dense case, loops of a given (say black) color 
are not constrained to visit all vertices.

The partition function of the fully-packed loop model is then defined by
assigning a weight $n_1$ per black loop and $n_2$ per white one, 
\eqn\pfpl{ Z_{\rm FPL}(n_1,n_2)= \sum_{{\rm fully-packed}\ {\rm loop}\atop
{\rm configurations} } n_1^{L_1} \, n_2^{L_2}  \ , }
where we have denoted by $L_i$ the total numbers of loops of each color $i=1,2$. 
Following~\JACO\ we shall denote the model with partition function \pfpl\ 
as the FPL$^2(n_1,n_2)$ model, while the densely packed version is referred to
as the DPL$^2(n_1,n_2)$ model.

The loop weights $n_i$ may be recast as local Boltzmann weights as follows. This step is important in the field theoretic description of the model,
since it leads to a {\it local} field theory. 
Let us assign to each black or white loop an arbitrary orientation, and attach to 
each vertex a local Boltzmann weight $e^{{\rm i} \pi (\epsilon_1 e_1+\epsilon_2 e_2)/4}$ where 
$\epsilon_i=1$ if the oriented loop of color $i$ makes a left turn, $\epsilon_i=0$ if
it goes straight, and $\epsilon_i=-1$ if it makes a right turn. Summing over all possible
orientations of all loops, we get a factor $2\cos\, \pi e_i$ per loop of color $i$,
and therefore we reproduce the desired loop weights by setting
\eqn\setting{ n_1=2\cos \, \pi e_1\ , \qquad n_2=2 \cos\, \pi e_2 \ . }

\subsec{Conformal field theory description}

\fig{A typical configuration of the FPL$^2$ model together with the 
bicoloration of its vertices (checkerboard of filled ($\bullet$) and empty ($\circ$)
dots). We have added the corresponding dictionary that allows to map
the loop configurations onto $A,B,C,D$ labelings of the edges.}{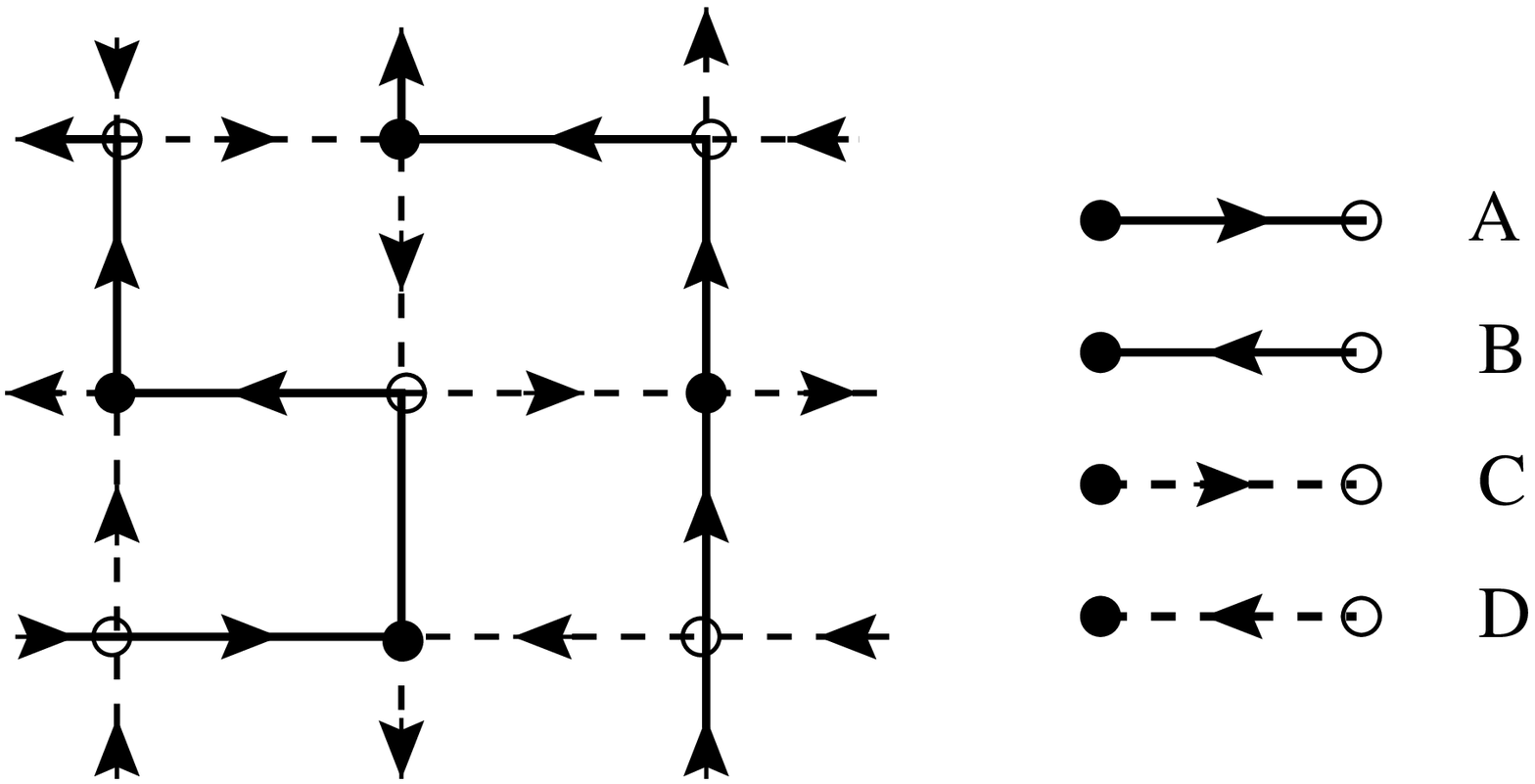}{8.cm}
\figlabel\dictio

The FPL$^2(n_1,n_2)$ model is known to be critical for $0\leq n_i \leq 2$~\JACO,
and is described in the continuum limit 
by a simple conformal field theory based on free scalar fields. 
To identify its basic degrees
of freedom, it is useful to rephrase the model as a (three-dimensional) 
height model as follows.
Starting from an oriented fully-packed black and white loop configuration, 
we first bicolor the vertices
of the square lattice, say with alternating filled ($\bullet$) and empty ($\circ$) dots.
Then we use the dictionary of Fig.~\dictio\
to assign one of the four labels $A,B,C,D$ to each colored and oriented edge. 
With this convention, it is clear that edges of type $ABAB\ldots$ alternate 
along black loops, whereas
edges of type $CDCD\ldots$ alternate along white loops, and that each vertex 
has one incident edge
of each type $A,B,C$ and $D$. It is seen that the four-labeling with $A,B,C,D$ is in one-to-one
correspondence with the coloring {\it and} orientation of edges of the FPL$^2$ model.
In particular, the orientation of a given black or white loop is reversed if we interchange
the $A$ and $B$ or $C$ and $D$ labels along the loop.

\fig{Rules determining the change of the height variable across labeled edges.
We adopt the Amp\`ere convention that the height is increased (resp.~decreased)
by the edge value if the arrow of the edge points to the left (resp.~right).
The edge labels must be interpreted as three-dimensional vectors with the 
respective values $\bf A$, $-\bf B$, $\bf C$, $-\bf D$.}{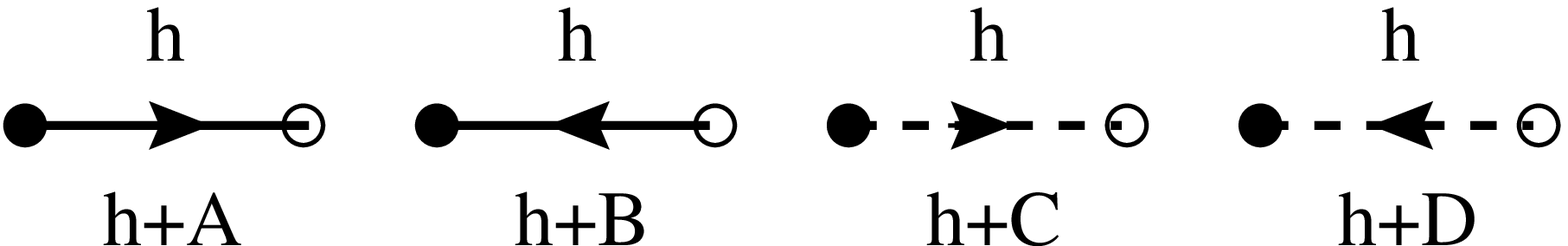}{8.cm}
\figlabel\ampere

The above colors allow for the definition of a dual vector height variable
on the center of each face of the lattice. Indeed, viewing as vectors the 
$A,B,C,D$ labeling of the edges
of the lattice, let us arbitrarily fix the height to be zero
on a given face of the lattice, and define it on all other faces by successive 
use the rules of Fig.~\ampere\ 
for the transition from a face to any of its neighbors.
Note that it is necessary to impose the condition ${\bf A}+{\bf B}
+{\bf C}+{\bf D}=0$ to
ensure that the heights are consistently defined around each vertex.
We may therefore assume in all generality that 
${\bf A},{\bf B},{\bf C},{\bf D}$ 
are actually four vectors in $\IR^3$
with vanishing sum. To get a more symmetric formulation, 
we may further fix 
${\bf A},{\bf B},{\bf C},{\bf D}$ 
to be the four unit vectors pointing from
the center of a tetrahedron towards its vertices.
The heights are then clearly three-dimensional, as linear combinations of 
${\bf A},{\bf B},{\bf C},{\bf D}$. 
In the continuum limit,
it was argued~\JACO\ that the three-dimensional height 
variable turns into a three-dimensional scalar
field. Moreover the symmetries of the model 
completely fix the action for these fields and the
corresponding field theory is conformal, with central charge
\eqn\centwo{ c_{\rm FPL}(n_1,n_2)= 3 - 6 \left( {e_1^2\over 1-e_1} + {e_2^2\over 1-e_2} \right)\ , }
where $e_i$ have been defined in \setting\ and are constrained by
$0\leq e_i \leq 1/2$.
Actually the shift in the central charge away from 3 is due to the introduction
of a background electric charge, ensuring that loops that have non-trivial
winding with respect to the periodic boundary conditions still get correctly
weighted, although for such loops the argument given before~\setting\ no longer
holds true.

\subsec{Meanders: the coupling to gravity}

To finally get to meanders, we must consider the coupling of the FPL$^2(n_1,n_2)$ model to
two-dimensional quantum gravity, by allowing the square lattice to fluctuate into arbitrary
planar four-valent graphs. 
For each such graph the fully-packed loop model is still defined
by coloring the edges black or white and allowing 
only the vertices shown in Fig.~\twovert. As before, each colored loop is
weighted by the appropriate $n_i$ factor ($i=1,2$). 

If we try to go through the steps of the previous section, namely by transforming the model
into a height model, the issue of bicolorability of the vertices of the lattice becomes crucial
on a random four-valent graph. 
Indeed, not all such graphs are vertex-bicolorable. So the coupling
to gravity {\it stricto sensu} (sum over arbitrary planar four-valent graphs) will destroy this property. 

We may now follow either of the two following paths. 
First, we can repair this and {\it impose}
that the particular coupling to gravity preserve the bicolorability, 
namely that the gravitational
model be defined on the set of vertex-bicolorable four-valent graphs only. 
These graphs are dual to the so-called  Eulerian quadrangulations. 
In genus zero (planar case), the latter are characterized by the fact that all their
vertices have an even valency (Euler condition). 
If we couple the FPL$^2(n_1,n_2)$ model to Eulerian gravity, the 
$A,B,C,D$ labeling is still well-defined, as well as the three-dimensional height,
now defined on the centers of the faces of the graph. 
This preserves the degrees of freedom of the flat space model entirely in the gravitational
formulation. This approach was initiated in \EKN\ for the simpler case
of the fully packed O$(n)$ model with only one type of loops.

\fig{Rules determining the change of the height variable across labeled edges
in the non-bicolored case.
These rules are identical to those of Fig.~\ampere, with the further restriction
that ${\bf B}=-\bf A$ and ${\bf D}=-\bf C$, allowing to ignore the 
bicoloration of vertices.}{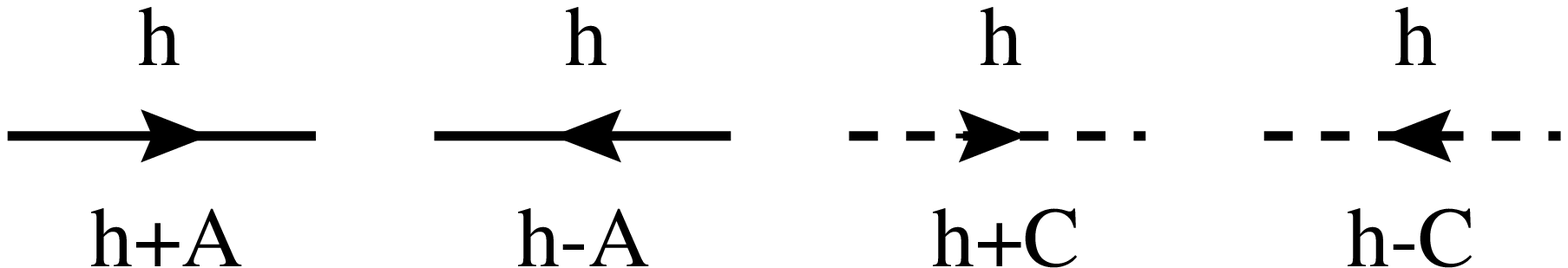}{8.cm}
\figlabel\twodim

On the other hand, we can study how the degrees of freedom of the model are affected by the
coupling to ordinary (non-Eulerian) quantum gravity. 
Having lost the bicolorability of vertices,
it is no longer possible to distinguish between $A$ and $B$ labels 
on one hand, and $C$ and $D$ 
on the other. We may still define an edge-labeling 
of the graph in one-to-one correspondence
with oriented colored fully-packed loop configurations on the graph, 
but with vectors
${\bf A},{\bf B},{\bf C},{\bf D}$
satisfying the two
constraints ${\bf A}+{\bf B}=0$ and ${\bf C}+{\bf D}=0$, 
and picking $\bf A$ and $\bf C$ to be two perpendicular 
unit vectors in $\IR^2$.  The correspondence between color/orientation and $A,C$ labels
reads as in Fig.~\twodim. As in Eq.~\pfpl, the model is further 
completed by attaching weights $n_i$ to each loop of color $i=1,2$.
We may now define a height variable on the centers of the faces of the graph, by use of the 
previous rules. The main difference is that the height now lives in two dimensions (the plane
generated by $\bf A$ and $\bf C$).
Such a dimensional reduction is also observed on the square lattice when going from the
FPL$^2(n_1,n_2)$ to the DPL$^2(n_1,n_2)$ model~\JACO; interestingly enough, the same 
dimensional reduction is also observed when the FPL$^2$ model is defined on
the Manhattan square lattice, with oriented loops respecting the Manhattan
orientation~\progress.
It results in a shift $c\to c-1$ in the central charge of the underlying conformal
theory, namely 
\eqn\newchar{ c(n_1,n_2)=c_{\rm DPL}(n_1,n_2)\equiv
2-6 \left( {e_1^2\over 1-e_1} + {e_2^2\over 1-e_2} \right)\ .}
In the following, we will concentrate on this formulation, eventually leading
to the solution of the meander problem.

The partition function of the fully-packed model coupled to {\it ordinary}
quantum gravity, hereafter referred to as the GFPL$^2(n_1,n_2)$ model, reads 
in genus zero:
\eqn\pfgra{ Z_{\rm GFPL}(n_1,n_2;x,y)=\sum_{{\rm four}-{\rm valent}
\ {\rm planar}\atop
{\rm graphs}\  \Gamma }
\sum_{{\rm FPL}\ {\rm configs. C}\atop
{\rm on } \ \Gamma}  
{1\over |{\rm Aut}(\Gamma,{\rm C})|} 
n_1^{L_1} n_2^{L_2} x^{V_a(\Gamma)} y^{V_b(\Gamma)}\ ,} 
where the sum extends over all the genus zero four-valent graphs $\Gamma$, and
$|$Aut$(\Gamma,{\rm C})|$ is the order of the 
symmetry group of $\Gamma$ equipped with the loop configuration C. 
We have also denoted by $V_a,V_b$ 
the total numbers of vertices of type a and b defined in Fig.~\twovert\
in the particular loop configuration, namely we have weighted each crossing of a 
black and a white  
loop by $x$ and each avoiding by $y$. When $x=y$, these are interpreted as the cosmological
constant, as the total number of vertices $V_a+V_b$ of $\Gamma$ is also the area 
of the corresponding
dual random surface.

\fig{Examples of (a) a tangent meander with 2 tangency points (b-vertices)
and 6 bridges (a-vertices) and (b) a meander with 8 bridges. 
}{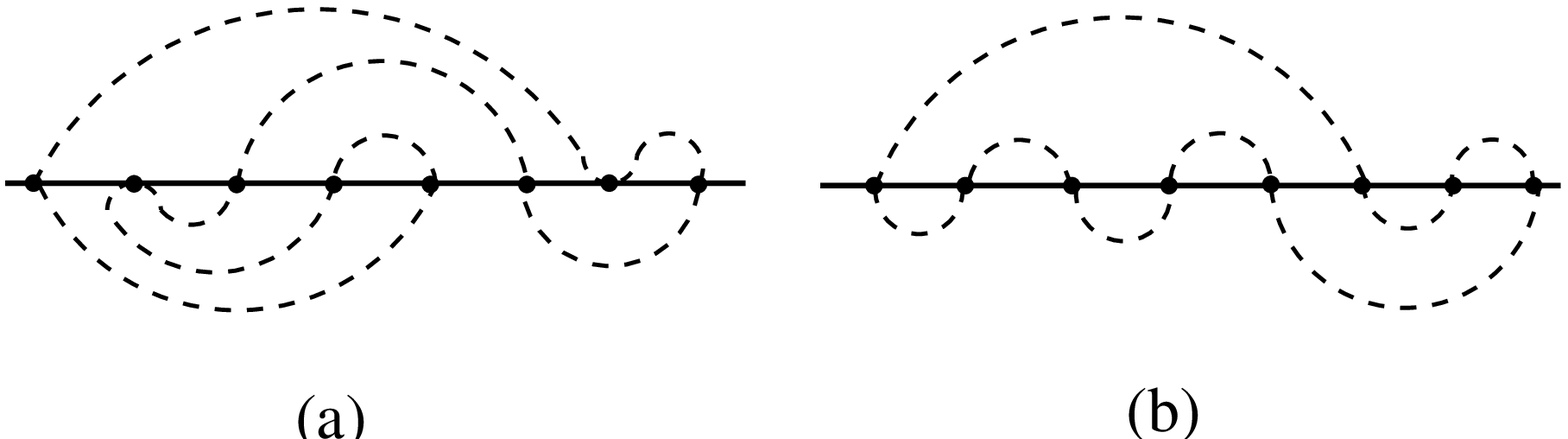}{12.cm}
\figlabel\tangent

To generate meanders, we must now extract from \pfgra\ the configurations with 
only one black and one
white loop, that will respectively play the role of the river and the road. 
This is done by taking the limit
$n_1,n_2\to 0$ in \pfgra, resulting in
\eqn\metan{\eqalign{ Z_{\rm GFPL}(x,y)&=\lim_{n_1,n_2\to 0}{1\over n_1n_2}
(Z_{\rm GFPL}(n_1,n_2;x,y)-1)\cr
&= \sum_{n,p \geq 0\atop
n+p\geq 1} {x^{2n} y^p\over 2(2n+p)} \mu_{2n,p}, \cr}}  
where we have denoted by $\mu_{2n,p}$ the total number of {\it tangent meanders} 
with $2n$ crossings
and $p$ tangency points, i.e. configurations of a non-selfintersecting circuit 
(road) crossing a line (river) through $2n$ points (bridges) and touching the 
river $p$ times (tangent contacts), as illustrated in Fig.~\tangent.
The usual meander numbers defined in \DGG\ correspond to only crossings
and no tangent points and read $M_n=\mu_{2n,0}$.
In \metan, the prefactor $1/(2(2n+p))$ stands for the symmetry factor attached 
to the tangent meanders:
going from the graph to the representation where the river is a line, we may 
indeed cut the river loop
in $2n+p$ places in between bridges and tangency points, and we still have 2 
choices for the up/down position.

\fig{A type b vertex of the FPL$^2$ gravitational model, together
with its dual height configuration. We note that the NE and SW heights
are identical. We may therefore undo the vertex as shown, which results in
its irrelevance.}{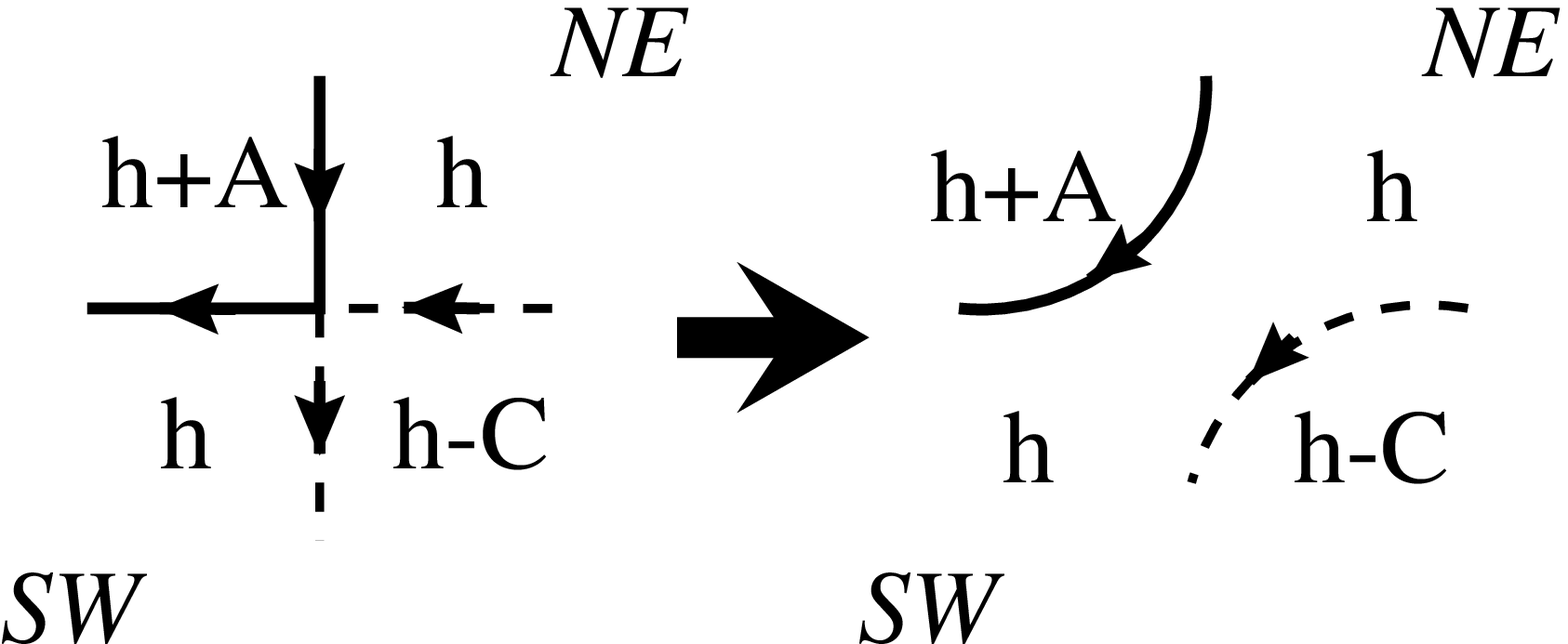}{6.cm}
\figlabel\irrel

The meanders are therefore generated by the function \metan\ for $y=0$. 
Let us now show that the 
universality class of the tangent meanders is the same as that of meanders. 
In the transformation into a (two-dimensional) height model, 
the ``tangency" vertex b of Fig.~\twovert\ 
corresponds to the arrangements of heights on adjacent faces
depicted in Fig.~\irrel.
We notice that the NE and SW heights are identical,
irrespectively of the orientations of the two loops. This means that as far as the height 
variable is concerned this vertex may be simply removed as shown. 
We conclude that the b-vertex 
of the model is irrelevant~\GGD. As a consequence, we expect the following asymptotics
for $N=2n+p$ large of the partition function $\mu_N(x,y)$
for tangent meanders with a total $N$ of bridges and tangency points:
\eqn\foas{ 
\mu_N(x,y)=\sum_{n, p \geq 0\atop 2n+p=N}x^{2n} y^p \mu_{2n,p}\sim 
C(x,y) { R(x,y)^N\over N^\alpha},}
where the configuration exponent $\alpha$ is independent of $x$ and $y$ for $x>0$ and
$y\geq 0$. In particular,
when $y=0$, $\alpha$ is identified with the
meander configuration exponent $M_n\sim c R^{n}/n^\alpha$.
The irrelevance of the vertex b will be checked numerically in subsection 4.4
below. 

More generally, we may
consider tangent meanders with one single river, but arbitrarily many possibly interlocking 
closed roads. This is given by the $n_1\to 0$ limit of \pfgra\ while $n_2=q$ is kept finite,
namely
\eqn\halpf{\eqalign{ Z_{\rm GFPL}(q;x,y)&=\lim_{n_1 \to 0}{1\over n_1}
(Z_{\rm GFPL}(n_1,n_2=q;x,y)-1) \cr
&=\sum_{n,p\geq 0; n+p\geq 1} {x^{2n} y^p \over 2(2n+p)} \mu_{2n,p}(q), \cr}}
where we have defined the tangent meander polynomial
\eqn\tanpol{ \mu_{2n,p}(q)=\sum_{k=1}^{n+p} \mu_{2n,p}^{(k)} q^k }
with coefficients $\mu_{2n,p}^{(k)}$ being the numbers of tangent meanders 
with $k$ connected components
of road, $2n$ bridges and $p$ tangency points. 
The polynomial $\mu_{2n,0}(q)=m_n(q)$ coincides with the meander polynomial defined in \DGG.
We may also define the canonical partition function:
\eqn\pfmu{ \mu_N(q;x,y)=\sum_{n, p \geq 0\atop 2n+p=N}x^{2n} y^p \mu_{2n,p}(q)
\ . }
We expect a large $N$ asymptotic behavior of the form
\eqn\larbeha{ \mu_N(q;x,y) \sim C(q;x,y) {R(q;x,y)^N \over N^{\alpha(q)}}, }
where the configuration exponent $\alpha(q)$ only depends 
on $q$ and not on $y/x$, due to the irrelevance of the b-vertex. 
In particular, it takes the same value at $y/x=0$, where
it coincides with the multi-component meander configuration exponent,
i.e. $m_n(q)\sim C(q) R(q)^{2n}/n^{\alpha(q)}$.

In the special case $q=1$ of arbitrarily many roads without extra weight, these numbers 
can be computed exactly.
Indeed, we may decompose an arbitrary multi-component tangent meander with $2n$ bridges
and $p$ tangency points into its upper part (above
the river) and lower part (below), and consider that these may be obtained first by
picking $p_1$ points among the total $N=2n+p$ to be the upper 
tangency points, and $p_2=p-p_1$ to
be the lower ones. Let us then draw small semi-circles 
tangent to the $p$ points, $p_1$ of them
in the upper half, $p_2$ in the lower. 
With the $2n$ bridges, we have now a total of $2n+2p_1$
points in the upper half and $2n+2p_2$ in the lower one to be connected among themselves 
by pairs through non-intersecting arches. Such upper and lower arch configurations have 
already been extensively studied in \DGG. In particular, there are $c_m=(2m)!/(m!(m+1)!)$
($c_m$ are the Catalan numbers) distinct upper arch configurations connecting $2m$ bridges 
by pairs in the upper half plane above the river.
Hence we must choose among the $c_{n+p_1}$ upper arch configurations 
and the $c_{n+p_2}$ lower ones
to form an arbitrary multi-component tangent meander.
This results in the formula
\eqn\forqone{ \mu_{2n,p}(q=1)= \sum_{p_1=0}^p 
{(2n+p)!\over p_1! (p-p_1)! (2n)!} c_{n+p_1} c_{n+p-p_1},}
where the combinatorial factor accounts for the choices of upper and lower 
tangent points among
the total of $2n+p$. The corresponding partition function \pfmu\ reads
\eqn\corpf{ \mu_N(q=1;x,y)=N! \sum_{n,p \atop 2n+p=N}{x^{2n}\over (2n)!} 
\sum_{p_1,p_2\atop p_1+p_2=p} 
{y^{p_1}\over p_1!} {y^{p_2}\over p_2!} c_{n+p_1} c_{n+p_2}. }
When $N$ is large, this is easily estimated by a saddle-point technique,
making use of the Stirling formula. For $x, y\geq 0$, we find the large $N$ behavior
\eqn\findthat{ \mu_N(q=1;x,y)\sim  {(4x+8y)^N \over N^3} }
up to a multiplicative constant depending on $x$ and $y$ only. 
This shows explicitly that the 
exponent $\alpha(q=1)=3$ is robust and is not 
affected by the respective values of $x$ and $y$.
This confirms in particular the above-mentioned irrelevance of the b-vertex, 
as $\alpha$ keeps to the same value, irrespectively of $y$.

In conclusion, the meander numbers belong to the universality class of the 
GFPL$^2(n_1,n_2)$ model at $n_1,n_2\to 0$. 
The corresponding flat space theory has 
the central charge \newchar\ with, according to \setting, $e_1=e_2=1/2$,
hence $c=-4$. 
In the next subsection, we will concentrate on the
partition function $Z_{\rm GFPL}(x,y=0)$ \metan\ that generates 
the meander numbers. 

More generally, the multi-component meander polynomial belongs to the universality
class of the GFPL$^2(n_1,n_2)$ model, with $n_1\to 0$ and $n_2=q$ and central charge
\newchar. 
Finally, we may also consider
meanders with arbitrarily many rivers and roads, with a weight $n_1$ per river and
$n_2$ per road; these objects belong to the 
universality class of the GFPL$^2(n_1,n_2)$ model. 
It has been shown using matrix model techniques that the GFPL$^2(n_1=1,n_2=q)$
with $y=0$ belongs
to the same universality class as the O$(q)$ model when coupled to ordinary
gravity (i.e. defined on arbitrary four-valent graphs) \CK . More precisely, the number 
of multi-river, multi-road meanders with a total of $2n$ intersections 
and with a weight $q$ per road and $1$ per river, also weighted by their inverse 
symmetry factor (and multiplied by $4n$ to make it comparable to $M_n$)
behaves for large $n$ as
\eqn\behoneq{ \eqalign{
C(1,q) &{R(1,q)^{2n}/n^{\alpha(1,q)}}, \qquad R(1,q)=2{\sin^2(\pi e/2)\over e^2},\cr
\alpha(1,q)&= {2-e \over 1-e}, \qquad q=2\cos(\pi e). \cr}}
The determination of $R(1,q)$ on which we have no prediction was made possible
by an explicit mapping of the matrix model onto a particular version of the
gravitational O$(q)$ model, where the critical value of the cosmological constant
$x_c$ can be explicitly calculated. 
The value $\alpha(1,q)$ 
corresponds to the expected scaling behavior of
a  $c=1-6 e^2/(1-e)$ (c.f. \newchar\ with $n_1=1$ and $n_2=q$)
conformal theory coupled to two-dimensional quantum gravity. 
In particular, when $q=1$, $c=0$ we get $\alpha(1,1)=5/2$ as expected in
``pure gravity" without matter. 

\fig{The transformation of the four-valent a- and b-vertices of the FPL$^2$ model
into pairs of trivalent ones connected by an extra (gray) edge, depicted as a
wavy line. 
Due to the 1 to 2 correspondence in the case of the a-vertex, we must
restrict ourselves to the case $x=2y$, and then each trivalent vertex receives a
weight $\sqrt{y}$. }{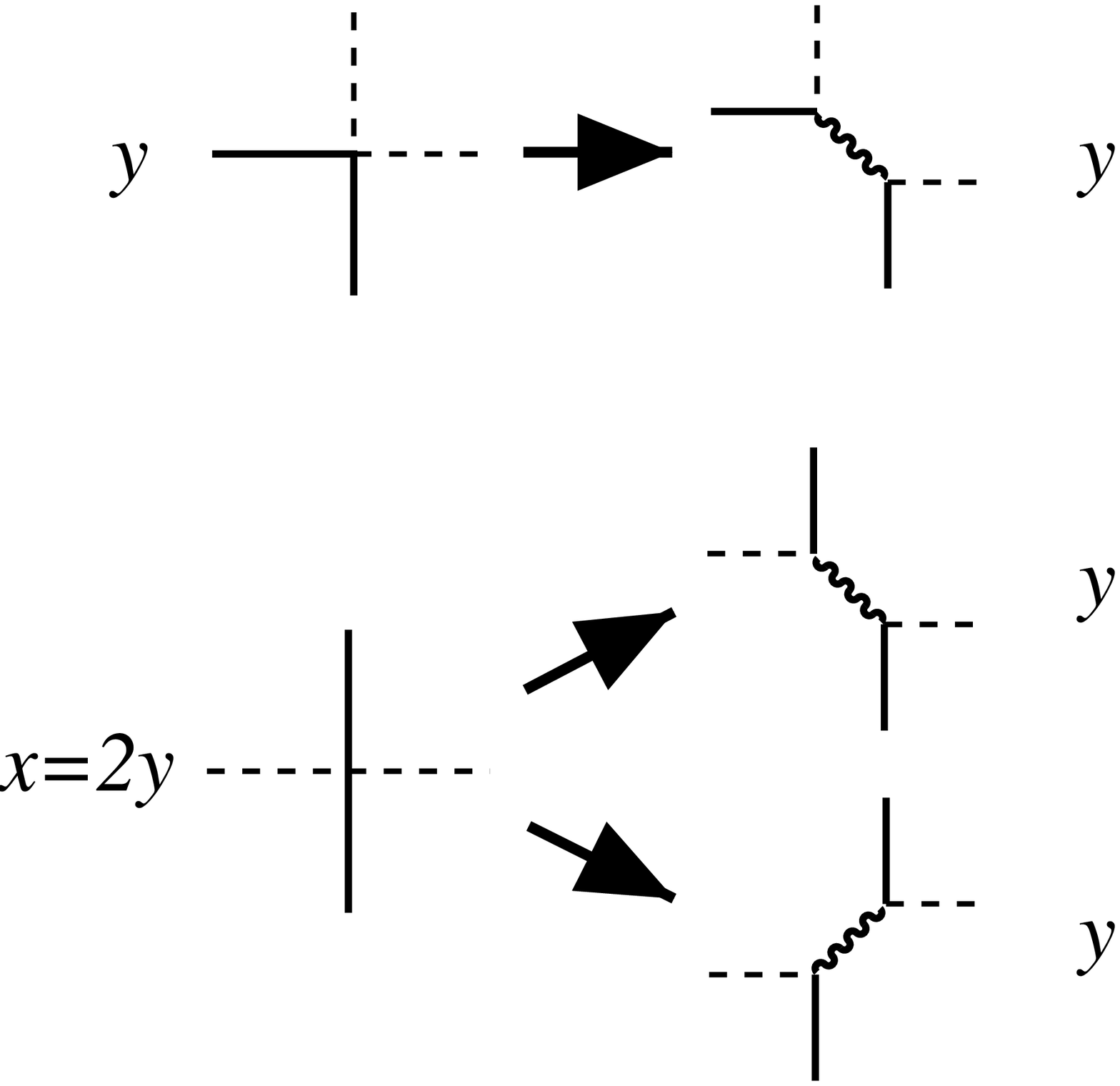}{5.cm}
\figlabel\trico

It is interesting to notice that if we restore the b-vertex (with a weight
$y$ per vertex), and choose specifically $x=2y$, then the GFPL$^2(1,q)$ model 
can be mapped onto that of tricoloring the edges of a random trivalent
graph (with black, gray and white edge colors) with a weight $q$ per 
loop of alternating say gray and white colors. 
Indeed, the quartic vertices a and b may be decomposed into pairs of connected
trivalent ones as shown in Fig.~\trico\
in which we have added a third type of edge, say with gray color.
This transformation allows to map the configurations of the GFPL$^2(1,q)$ model 
weighted by
$x=2y$ per a-vertex and $y$ per b-vertex onto those of arbitrary edge-tricolored  
trivalent graphs, with a weight $\sqrt{y}$ per vertex and $q$ per white/gray loop. 
The latter model 
was solved in \KE\ and identified with a particular version of the fully-packed
O$(2q)$ model
on random trivalent graphs: indeed, the white and gray edges form naturally
fully packed loops on the graph, and on each loop we may interchange the  
two colors to generate new tricolorings, hence we may simply draw fully-packed loops on 
trivalent graphs, and attach a weight $2q$ to each of these loops. But the loops have all
even lengths and this turns out to dimensionally reduce the model to an effective
O$(2q/2=q)$ one with central charge $c=1-6e^2/(1-e)$ as before. 
This particular point therefore lies in the same
universality class as the model without b-vertex, which gives another explicit example of
the irrelevance of the b-vertex.

\subsec{Field theory description of meandric numbers}

The coupling of a conformal field theory with central charge $c\leq 1$ to two-dimensional
quantum gravity (i.e. its definition on random surfaces)
has a simple field-theoretical formulation in terms of the Liouville
field describing the conformal classes of metrics of the surfaces. This has led to a
number of results, including the precise determination of various critical exponents. 
Indeed, the gravitational theory (say on genus zero surfaces) displays a critical
behavior as a function of the cosmological constant $x$. In particular, there exists a
finite value $x_c$ of $x$ at which the (connected) partition function behaves as
\eqn\behagra{ Z(x)\sim (x_c-x)^{2-\gamma(c)}, }
where the string susceptibility exponent $\gamma$ is related to the central charge $c$
through \KPZ\
\eqn\kpz{ \gamma(c)={c-1-\sqrt{(1-c)(25-c)} \over 12}. }
When applied to the DPL$^2(0,0)$ model of the previous section (with $c=-4$), 
whose gravitational version was shown to describe meanders, we find that
\eqn\gamame{ \gamma\equiv \gamma(-4)=-{5+\sqrt{145} \over 12}\ .}
Comparing \behagra\ with the expansion 
\eqn\mexp{ Z_{\rm GFPL}(x)=\sum_{n\geq 1} {x^{2n}\over 4n} M_n }
we deduce the asymptotic behavior \GGD 
\eqn\meas{ M_n \sim C {x_c^{-2n} \over n^\alpha}, 
\qquad \alpha=2-\gamma={29+\sqrt{145}\over 12}.}

Moreover, a number of the operators of the flat space conformal theory (in particular the
spinless ones, with conformal dimensions $h={\bar h}$) get dressed by
gravity, in such a way that they acquire anomalous scaling dimensions. Any given operator
$\phi_k$ with dimensions $h_k={\bar h}_k$, is dressed into an operator ${\tilde \phi}_k$
with dressed dimension $\Delta_k$ such that the correlation functions behave as
\eqn\behacorr{ \langle {\tilde \phi}_{k_1} {\tilde \phi}_{k_2} \ldots 
{\tilde \phi}_{k_p}\rangle
\sim (x_c-x)^{2-\gamma+\Sigma (\Delta_{k_i}-1)} }
when $x$ approaches the critical value $x_c$, and 
where the dressed dimension $\Delta_k$ is related to
the flat space conformal dimension $h_k$  through
\eqn\kpzdim{ \Delta_k={ \sqrt{1-c+24 h_k} -\sqrt{1-c} \over \sqrt{25-c}-\sqrt{1-c}}. }

Let us now present the operator content of the $c=-4$ 
conformal theory describing the dense loop model
DPL$^2(0,0)$~\JACO. 
For generic values of $n_1,n_2$, the DPL$^2(n_1,n_2)$
has a continuum description as a two-component scalar 
field with charges at infinity. More precisely, 
it is a Coulomb gas made of two decoupled scalar fields, 
with $c=c(n_i)=1-6 e_i^2/(1-e_i)=-2$ at $n_i=0$ ($e_i=1/2$) respectively, 
each viewed as the effective field theory of loops of one color. 
In particular, within each scalar field theory 
(indexed by the color $i=1,2$), there exist operators
$\psi_k^{(i)}(z)$ that create $k$ 
oriented defect lines (of color $i$) for the scalar field, 
with conformal dimensions \DaDu :
\eqn\dimpsi{ h_k^{(i)} = {k^2-4 \over 32} }
at $n_i=0$. In the Coulomb gas formalism, these correspond to electromagnetic
operators with electric charge $e_i$ (spin-wave) and magnetic charge $\pm k/2$
(vortex), according to
whether the defect line is oriented from or to the insertion point, and $k$ is
a strictly positive integer.
The electric charge ensures that, if the defect lines wind around the 
insertion point, all extra curvature weights get cancelled~\Nienh.
For $k=0$, \dimpsi\
must be replaced by $h_0^{(i)}=0$ corresponding to the identity operator.
The correlation functions must have a vanishing total magnetic charge.
Although these operators also carry an electric charge $1/2=e_1=e_2$, 
the electric neutrality of correlators imposes no extra 
condition\foot{ 
It can be shown \DSZ\ that the corresponding non-minimal
conformal theory pertaining to the loops of color $i$ (with central charge
$c=1-6e_i^2/(1-e_i)$) is equally well defined as a theory with central charge $c=1$,
in which the operator content has been reorganized, and in particular
all of the above electromagnetic
operators get a vanishing electric charge, while their magnetic one is unchanged.
A supersymmetric version of this theory was also introduced in \SAL, in which
arbitrary correlators could be calculated, irrespectively of the electric charge.
}.

We may also combine operators for both colors, namely consider mixed operators 
$\psi_{k_1,k_2}=\psi_{k_1}^{(1)}\psi_{k_2}^{(2)}$ with conformal dimension 
$h_{k_1,k_2}=h^{(1)}_{k_1}+h^{(2)}_{k_2}$ for $k_1,k_2\in \IZ$.

Let us now study the dressing of these operators by gravity, 
and interpret them in meandric terms.
The dressed operator ${\tilde \psi}_k^{(i)}$ again corresponds to the
creation of a vertex with $|k|$ outcoming ($k>0$) or incoming ($k<0$)
lines of color $i$.
 
As a first example, note that $h_2^{(i)}=0$. From the formula \behacorr, it is clear 
that the operators ${\tilde \psi}_{\pm 2}^{(i)}$
have the effect of {\it marking} a point on a loop of color $i$. 
However, due to the constraint of global magnetic neutrality of correlators,
the marking is not arbitrary as the loop of color $i$
must have an even number of
marks (alternating ${\tilde \psi}^{(i)}_2$ and ${\tilde \psi}^{(i)}_{-2}$).
In particular, the 
one-point function $\langle {\tilde \psi}_2^{(1)}\rangle$
that would naively count meanders with a marked river 
turns out to vanish, while for instance
the two-point function 
$\langle {\tilde \psi}_2^{(1)}{\tilde \psi}_{-2}^{(1)}\rangle$ 
indeed counts meanders with two marked points
on  the river. We will meet more examples of this below. 

\fig{A typical semi-meander configuration.}{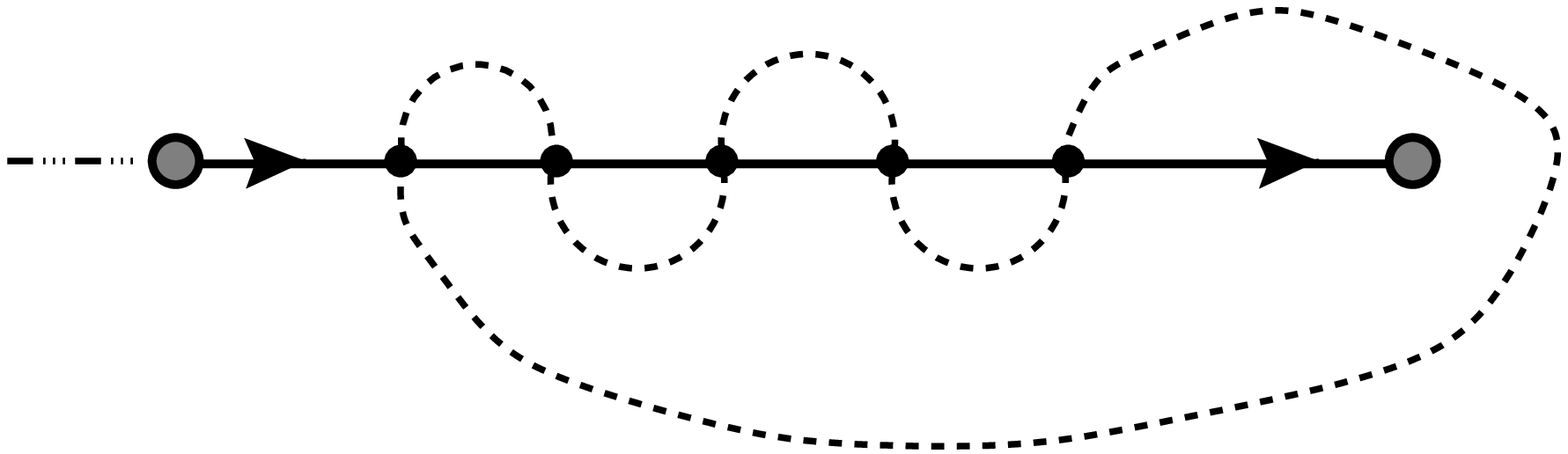}{7.cm}
\figlabel\semifig

The first application of the above concerns the two point function 
describing the insertion of a segment of river (color 1) 
\eqn\tpone{ \langle {\tilde \psi}_1^{(1)} {\tilde \psi}_{-1}^{(1)}\rangle 
\sim (x_c-x)^{2\Delta_1-\gamma}. }
Recall that when we take the limit $n_1,n_2\to 0$, only diagrams 
with one connected component
of river and one of road are selected. In the case of \tpone, 
the river forms a segment, around which
the road can freely wind (see Fig.~\semifig). 
To fix ambiguities, let us send one end of the river to infinity
(say to the left) and therefore represent 
the river as a half-line (this is allowed as
we work on Riemann sphere). 
The number of configurations 
of a closed road crossing a half-line (river with a source) 
through $n$ bridges is defined as the semi-meander number ${\bar M}_n$. 
We immediately identify the series expansion of \tpone\
as a function of $x$ to be
\eqn\serexP{ \langle {\tilde \psi}_1^{(1)} {\tilde \psi}_{-1}^{(1)}\rangle 
= \sum_{n\geq 1} {\bar M}_n x^n.}
We therefore deduce the semi-meander asymptotics \GGD
\eqn\semi{ {\bar M}_n\sim {\bar c} {{x_c}^{-n}\over n^{\bar \alpha}}
\qquad {\bar \alpha}=1+2 \Delta_1-\gamma= 1+{\sqrt{11}\over 24}(\sqrt{5}+\sqrt{29}). }

\fig{A typical example of an eight figure river geometry (a). On the
sphere, it is equivalent to the situation of (b), where the river crossing
has been cleared of all winding pieces of road. This crossing may finally be sent to
infinity (c) so as to form two parallel rivers. We have represented a particular
configuration of road with $2p_1=4$ bridges on one loop and $2p_2=2$ on the other.
Due to magnetic neutrality in the operator formulation, the two loops are
marked as shown.}{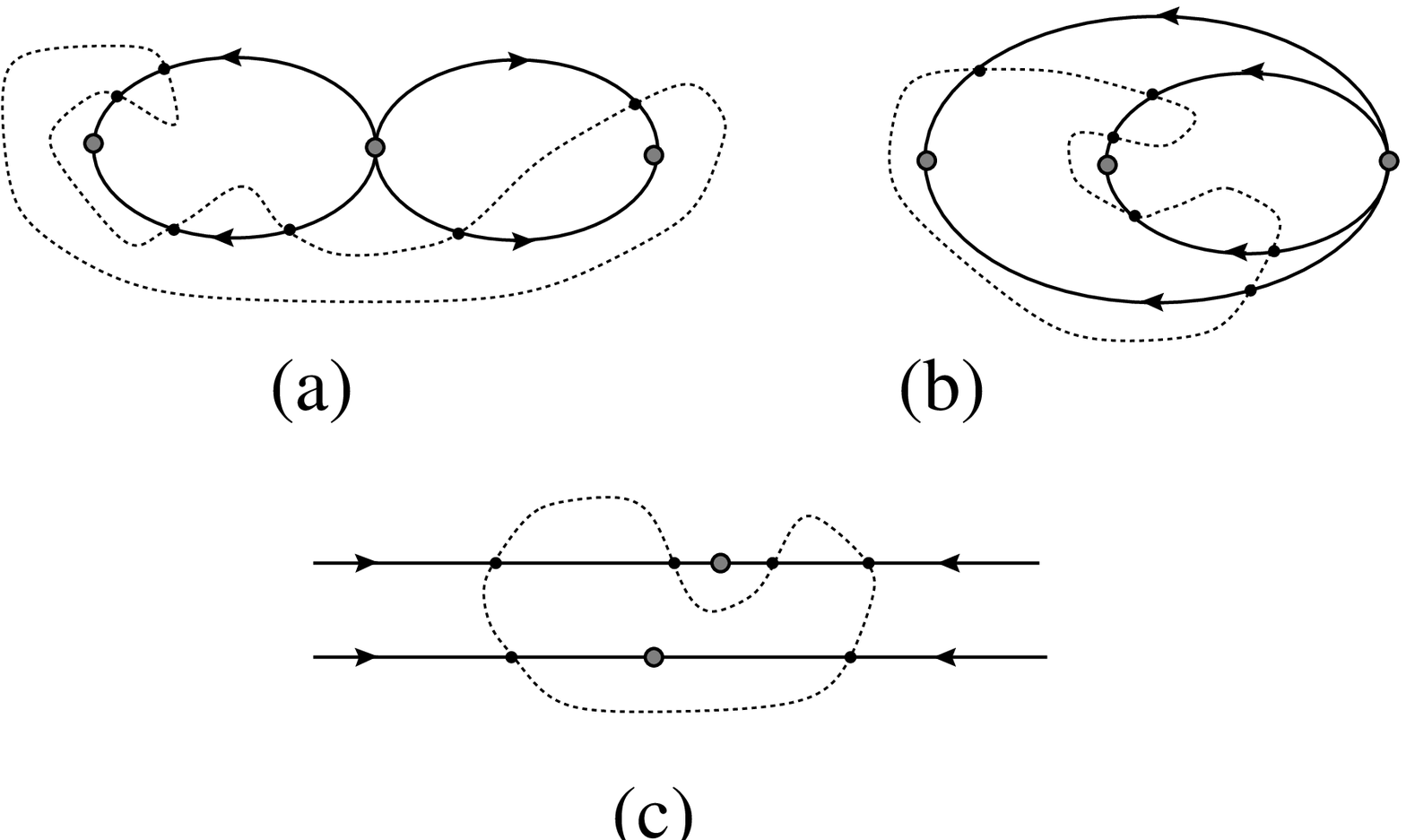}{10.cm}
\figlabel\eight

Our second example is the ``eight" meander, in which the river 
forms the figure ``eight". To generate
it, we need to insert an operator ${\tilde \psi}_4^{(1)}$, 
but magnetic neutrality forces us to
insert two river-marking operators ${\tilde \psi}_{-2}^{(1)}$, 
one on each loop of the eight.
The three-point correlation function $\langle {\tilde \psi}_4^{(1)} 
({\tilde \psi}_{-2}^{(1)})^2 \rangle$
therefore generates the numbers of meanders whose river forms an eight 
figure, and with one marked
point on each loop (see Fig.~\eight\ (a)). 
To fix ambiguities, let us send the river-crossing to 
infinity. We may now represent
the river as two parallel lines (all connected to the four-valent point 
at infinity), but the markings are still there (c.f. Fig.~\eight\ (c)). Hence we have
\eqn\intpt{\eqalign{ 
\langle {\tilde \psi}_4^{(1)} ({\tilde \psi}_{-2}^{(1)})^2 \rangle &= 
{1\over 2}\sum_{n\geq 1}
x^{2n} M_n^{\rm 2-mark} \cr
&={1\over 2}\sum_{n\geq 1}
x^{2n}\sum_{p_1+p_2=n} (2p_1+1) (2p_2+1) M_{p_1,p_2}, \cr} }
where $M_n^{\rm 2-mark}$ is the number of meanders with two rivers, 
each of which is marked,
and $M_{p_1,p_2}$ is the number of meanders with two parallel rivers, 
and with $2p_1$ bridges
one the first and $2p_2$ on the second (there are indeed $2p_i+1$ possible markings on a
river loop with $2p_i$ bridges). 
Note the prefactor of $1/2$ accounting for the fact that we distinguish the top and bottom
of the figure when we represent a two-river meander.
From \behacorr, we find that \GGD
\eqn\marqme{\eqalign{
M_n^{\rm 2-mark} &\sim {\rm const.} {x_c^{-2n} \over n^{\alpha_{\rm 2-mark}}}, \cr
\alpha_{\rm 2-mark}&=\Delta_4+2\Delta_2-\gamma= 
{1\over 24}(\sqrt{5}+\sqrt{14})(\sqrt{5}+\sqrt{29}). \cr}}

\fig{A meander with two parallel rivers may be viewed as a meander
``with a seam". The point where the seam is attached is a remnant of the
point at infinity (say to the left), where the two rivers meet. The seam
just prevents roads from encircling this point.}{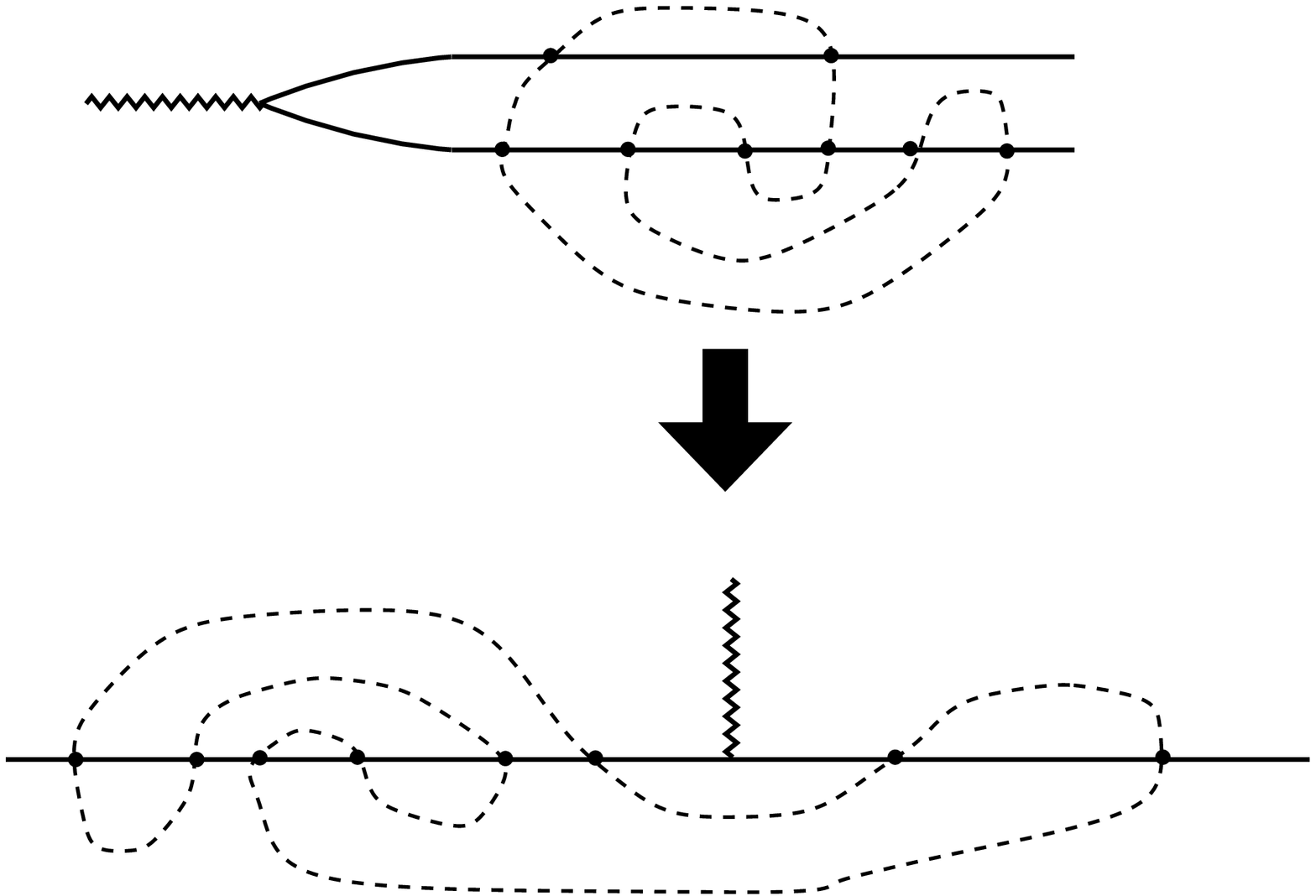}{6.cm}
\figlabel\seam

It is also easy to determine the large $n$ behavior of the unmarked meanders with two
parallel rivers. Let us draw the two rivers horizontally. We may view the two
rivers as connected to each other at their far left (at infinity) and represent them
as in Fig.~\seam\ as one marked line, such that a seam originates from the marked point
that prevents any road from encircling the point to the left. So we have mapped two-river
meanders onto one-river meanders with a seam. We may now generate all meanders with a seam
by considering all the ways of placing a seam on the upper-half of meanders. If $E(M)$
denotes the total number of upper exterior arches of a meander $M$ (namely arches that
have no other arch above them), then there are $E(M)+1$ distinct ways of decorating
the meander with a seam. Hence the total number of two-river meanders reads
\eqn\tottworiv{ M_n^{\rm 2-riv.}= \sum_{{\rm meanders}\ M\atop
{\rm with}\ 2n \ {\rm bridges}} (E(M)+1) = \langle E+1 \rangle_n M_n, }
where $\langle \ldots \rangle_n$ stands for the average over all meanders with
$2n$ bridges.
\fig{How to construct a new meander with $2n+2$ bridges 
from a meander with $2n$ bridges
and one distinguished exterior arch, here on the upper half of the meander. 
We cut the corresponding arch, and
glue it back across the river, by encircling the whole lower part of 
the meander. This is easily inverted within the set of meanders with
$2n+2$ bridges and one unique upper or lower exterior arch, showing
that this construction gives only rise to distinct meanders.}{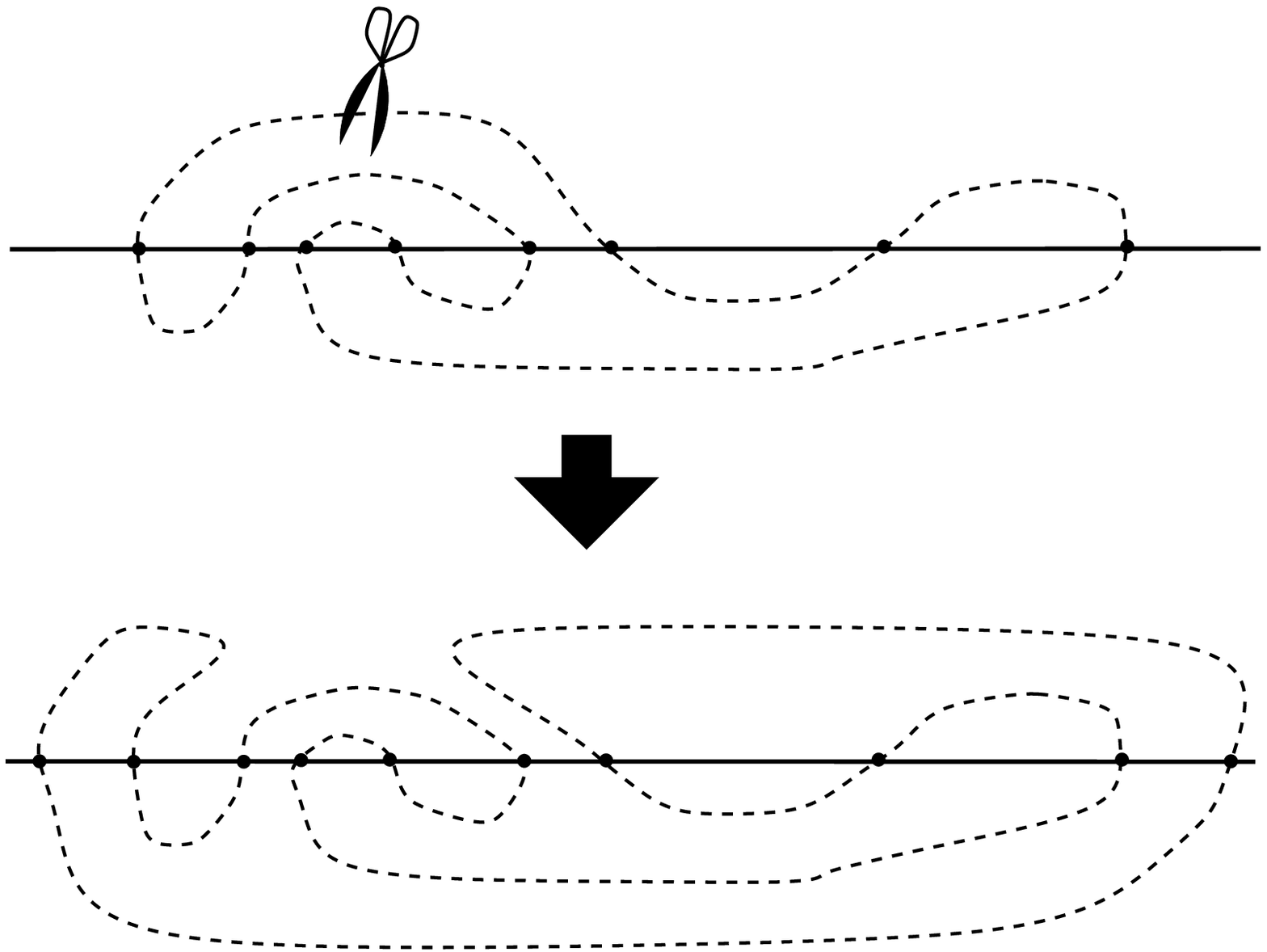}{6.cm}
\figlabel\bound
We now argue that $\langle E+1 \rangle_n$ remains finite when $n\to \infty$.
Indeed, starting from a meander with $2n$ bridges, we may construct one new meander
with $2n+2$ bridges for each upper and lower exterior arch. As shown in Fig.~\bound, we
simply cut the arch and close it on the other half, thus creating two new bridges
and keeping a single connected road. Such meanders have $2n+2$ bridges and one single
exterior arch on the side where we have closed the arch, and are all distinct.
Hence we may write that their number is bounded by $M_{n+1}$, namely
\eqn\bounda{ 2 \sum_{{\rm meanders}\ M \atop
{\rm with}\ 2n \ {\rm bridges}} E(M) \leq M_{n+1}\ \ \  \Rightarrow \ \ \ 
\langle E \rangle_n \leq  {1\over 2} {M_{n+1}\over M_n} \to {R \over 2}. }
We therefore have the asymptotic bounds 
$1\leq \langle E+1\rangle_n \leq 1+R/2$, which implies that $M_n^{\rm 2-riv.}\propto
R^{2n}/n^\alpha$ like ordinary meanders, hence $\alpha_{\rm 2-riv.}=\alpha$.
We also deduce that
\eqn\limnonetwo{\eqalign{
&\langle (2p_1+1)(2p_2+1)\rangle_n\equiv
{\sum_{p_1+p_2=n} (2p_1+1) (2p_2+1) M_{p_1,p_2} \over \sum_{p_1+p_2=n} M_{p_1,p_2}}
\sim n^\beta\ , \cr
&\beta= \alpha -\alpha_{\rm 2-mark} =2-\Delta_4=2-{1\over
24}(\sqrt{14}-\sqrt{5})(\sqrt{5}+\sqrt{29})=1.521898\cdots. \cr}}
This shows the rather unexpected result that the two-river meanders tend to be
very asymmetric, with a number of bridges of the order $n$ on one river
(since $p_1+p_2=n$) and
of the order $n^{\beta-1}$ on the other, with $\beta-1=0.521898\cdots$. 
The above arguments can be generalized to the case of multi-component meanders
and give rise to similar predictions.

In \GGD, a number of other results have been presented, 
all corresponding to more sophisticated
river geometries, and making use of the magnetic defect operators ${\tilde \psi}_k^{(i)}$.  
It would also be possible in principle to use mixed operators to generate diagrams
with both road and river geometries fixed.
Another direction consists in going away from the point $n_1=n_2=0$, for instance by
considering the GFPL$^2(n_1=0,n_2=q)$ model in which meanders 
with arbitrary numbers of roads
are considered, with a weight $q$ per road. In that case, the 
corresponding conformal theory
with central charge \newchar\
\eqn\charge{ c= -1 -6 {e^2 \over 1-e} \qquad e={1\over \pi}{\rm Arccos}({q\over 2}) }
may be viewed as two decoupled bosonic field theories: a $c=-2$ theory 
(that of river loops, at $n_1=0$) 
and one with $c=c(q)=1-6 e^2/(1-e)$, where $q=2 \cos \pi e$ (that of road loops). 
The generating function $Z_{\rm GFPL}(q;x,y=0)$ \halpf\ for meander polynomials reads
\eqn\mepof{
Z_{\rm GFPL}(q;x,y=0)=\sum_{n\geq 1} {x^{2n}\over 4n} m_n(q) \sim(x_c-x)^{2-\gamma(c)} }
with $\gamma(c)$ as in \kpz, and $c$ as in \charge.
This turns into the asymptotics
\eqn\asmq{  m_n(q) \sim C(q) {R(q)^{2n} \over n^{\alpha(q)}}, 
\qquad \alpha(q)=2-\gamma(c), }
namely \GGD
\eqn\almeq{ \alpha(q)= 2+{1-e+3e^2+\sqrt{(1-e+3e^2)(13-13e+3e^2)}\over 6(1-e)} }
with $e$ as in \charge.
An analogous formula holds for the (multi-component) semi-meander polynomials, namely
\eqn\semqpo{ {\bar m}_n(q) \sim {\bar C}(q) {{\bar R(q)}^n \over n^{\bar \alpha(q)}},
\qquad {\bar \alpha}(q)= 1+2 \Delta_1(c) -\gamma(c) }
with $\Delta_1(c)$ as in \kpzdim, $c$ as in \charge, and $h_1=-3/32$ as in \dimpsi.
This yields \GGD
\eqn\alsm{ {\bar \alpha}(q)=1+{\sqrt{2(24e^2+e-1)}
(\sqrt{1-e+3e^2}+\sqrt{13-13e+3e^2})\over
24(1-e)} }
with $e$ as in \charge.

An important remark is in order concerning the range of validity of \almeq\ and \alsm.
First, the DPL$^2(0,q)$ model is critical only for $q \leq 2$, i.e.
$e \geq 0 $.
We expect therefore a very different
scaling behavior for the meandric numbers when $q>2$. At large $q$, it
was shown in Ref.~\NOUS\ that 
\eqn\abovone{ \alpha(q)={3 \over 2}}
independently of $q$.

\fig{A typical ``branched" semi-meander.}{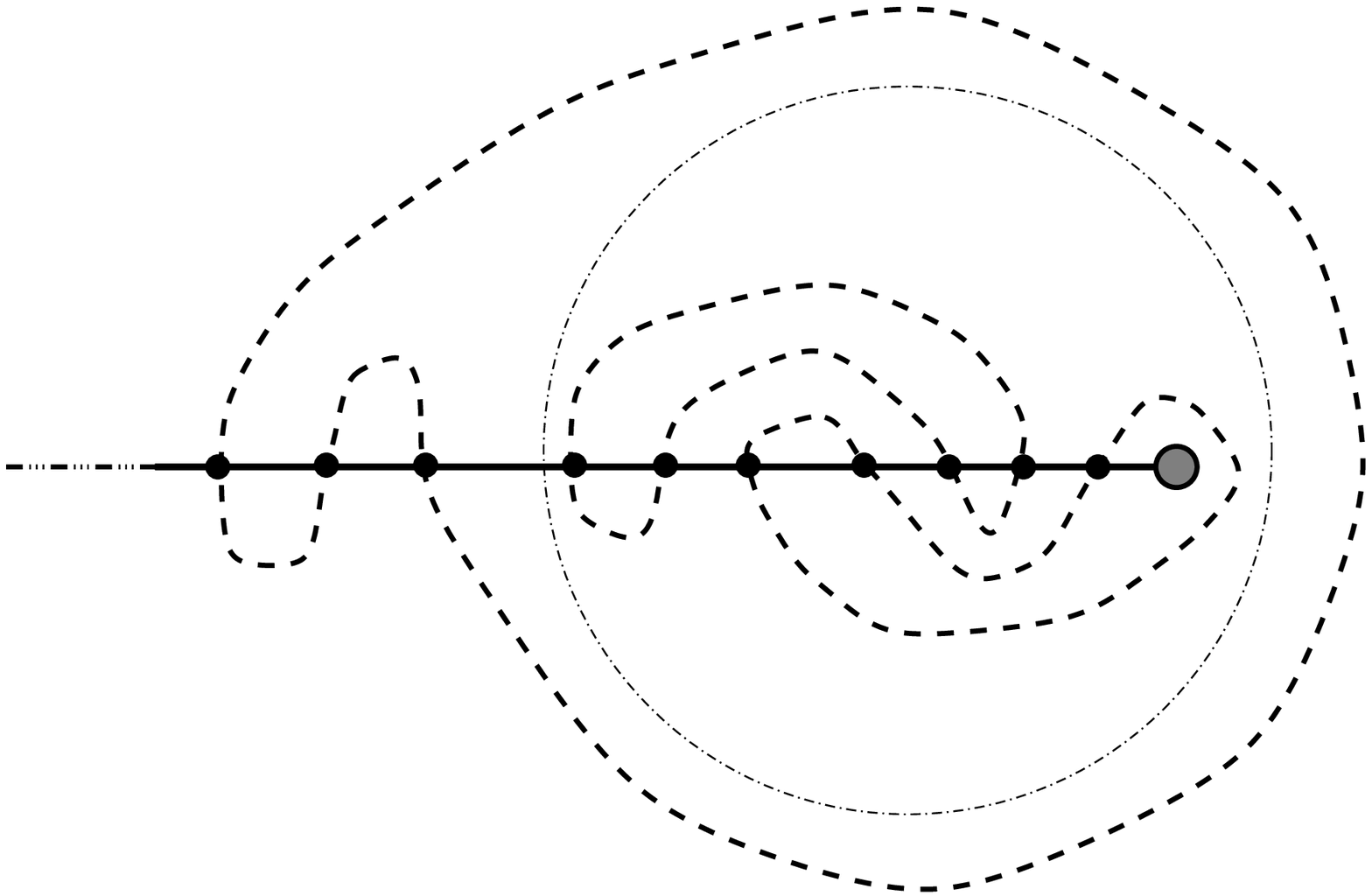}{6.cm}
\figlabel\bigcirc

On the other hand, for semi-meanders, this transition at $q=2$ is never 
reached. Indeed, another
phenomenon appears at a value $q_{\rm c}<2$. 
It corresponds to a proliferation of connected components of road and also of the pieces
of road that wind around the source of the river. 
More precisely, let us
compare the two following contributions to the multi-component semi-meander numbers,
according to whether:

\item{(i)} ``branched" semi-meanders dominate: the semi-meander is typically separated into 
two parts, namely the interior and exterior of a ``big circle" as shown in Fig.~\bigcirc.
The contribution of such semi-meanders reads asymptotically
\eqn\contcirc{ \sum_{n_1+n_2=n}{{\bar R}(q)^{n_1}\over n_1^{{\bar \alpha}(q)}} \times 
{{\bar R}(q)^{n_2}\over n_2^{{\bar \alpha}(q)}} \sim 
{{\bar R}(q)^{n}\over n^{2{\bar \alpha}(q)-1}}. }

\item{(ii)} ``connected" semi-meanders dominate: 
the semi-meanders cannot be cut as in (i), and
typically behave as 
\eqn\bm{ {{\bar R}(q)^{n}\over n^{{\bar \alpha}(q)}}. }

The transition between the two regimes (i)-(ii) will take place when $2{\bar
\alpha}(q)-1={\bar \alpha}(q)$.
We deduce that precisely at the transition, we must have $\bar\alpha=1$, which according
to \alsm\ takes place when $24 e^2+e-1=0$, namely at the critical value $e=e_{\rm c}$, $q=q_{\rm c}$
given by
\eqn\ecrit{ e_{\rm c}={\sqrt{97}-1 \over 48}, \qquad q_{\rm c}=2 \cos \, \pi {\sqrt{97}-1 \over 48}, }
also corresponding to $c=3/4$.
Hence the formula \alsm\ is only valid for $e_{\rm c}\leq e$, namely 
$q\leq q_{\rm c}=1.673849\cdots<2$.
Beyond $q=q_{\rm c}$, we expect the average number of connected components of 
road to be of the order of $n$, and we have \NOUS\
\eqn\abotrans{ {\bar \alpha}(q)= 0 \qquad {\rm for } \ \ \  q>q_{\rm c}. }

It is quite interesting to notice that the transition of semi-meanders is of a 
different nature than that of meanders. The latter simply encounters the
$q=2$ transition of the O$(q)$ model.
The former undergoes a winding transition, in which semi-meanders become very different
from meanders. Indeed, as long as the dominant semi-meanders have very little 
winding around the
source of the river, they behave just like meanders (hence $R(q)={\bar R}(q)$ in the
regime $q<q_{\rm c}$). But when the winding number of the semi-meanders becomes relevant (of
the order of $n$) we expect many more semi-meanders than meanders with the same number of
bridges, and ${\bar R}(q)>R(q)$ for $q>q_{\rm c}$.

\newsec{Meander enumeration algorithms}

In this Section, we will describe the algorithms that we used
to check numerically our predictions of Sect.~2 for the different
configuration exponents. All the results presented
below concern the case $n_1=0$, i.e. a connected river configuration, and
a varying weight $n_2=q$ per road, i.e. multi-connected road configurations. 
We will also consider the several meander geometries discussed in Sect.~2. 

All our algorithms are based on transfer matrix techniques.
Transfer matrices have proven to be very powerful for studying a wide 
range of statistical mechanics systems, especially in two dimensions. 
Originally applied to systems with local interaction, such as the Ising
model, the ``row to row" transfer matrix describes the transition
between two successive rows of spins, say at `time' $t$ and $t+1$ along the 
transfer direction. More recently, the transfer matrix technique
was also applied to systems with  non-local degrees of freedom, but where,
in a suitably designed basis, the statistical weights can still be
evaluated between neighboring time slices.
In this way it became possible to study self-avoiding polygons \Enting\ and 
walks \Derrida, and the random cluster model \Blote, to mention but a few
important examples.

Finally, transfer matrices can also be applied to random
lattice problems, as long as a definite transfer direction can be defined,
as for instance in the case of Lorentzian gravity \LOR .
This is also precisely the case in the meander problem, where we
can simply choose to transfer along the river, adding one bridge in each
time step.  This was first 
recognized by Jensen in \JEN\ where the method proved to be much better 
than previous enumeration algorithms, leading in particular to the 
largest accessible numbers of bridges. 

When implemented on a computer, the transfer matrix algorithms will allow
to enumerate {\it exactly} the various meandric objects that we are 
interested in for a fixed finite number $N$ of bridges up to a certain 
maximal value of $N$ (typically up to 48 bridges in the results presented
below).  
From these exact finite $N$ values, we can extract estimates for
the large $N$ asymptotic behaviors, and in particular for the configuration 
exponents described in Sect.~2. We can then compare these estimates
to the corresponding expected theoretical values.

This section is organized as follows: The case of one infinite river 
(the original meander problem) is discussed 
in subsection 3.1, while the other geometries, including the case of two 
infinite parallel rivers (connected at infinity), with possibly marked points 
on each of them, and the case of a semi-infinite river (semi-meanders), 
are discussed in subsection 3.2. In the three cases above, we allow
only crossing vertices of type a (cf.~Fig.~\twovert).
The inclusion of tangency points for
the infinite river case (tangent meanders) is discussed in subsection 3.3.

\subsec{One infinite river (meanders)}

\fig{A multi-road meander with 8 bridges and
 2 connected components. Upon addition of a new bridge, four different
 operations (O, C, U and D) are allowed, as discussed in the text. We also
 show the state of that part of the meander that is to the left of a given
 vertical line perpendicular to the river (here in position after $t=2$
steps). The transfer matrix acts by transferring this line from the left
to the right.}{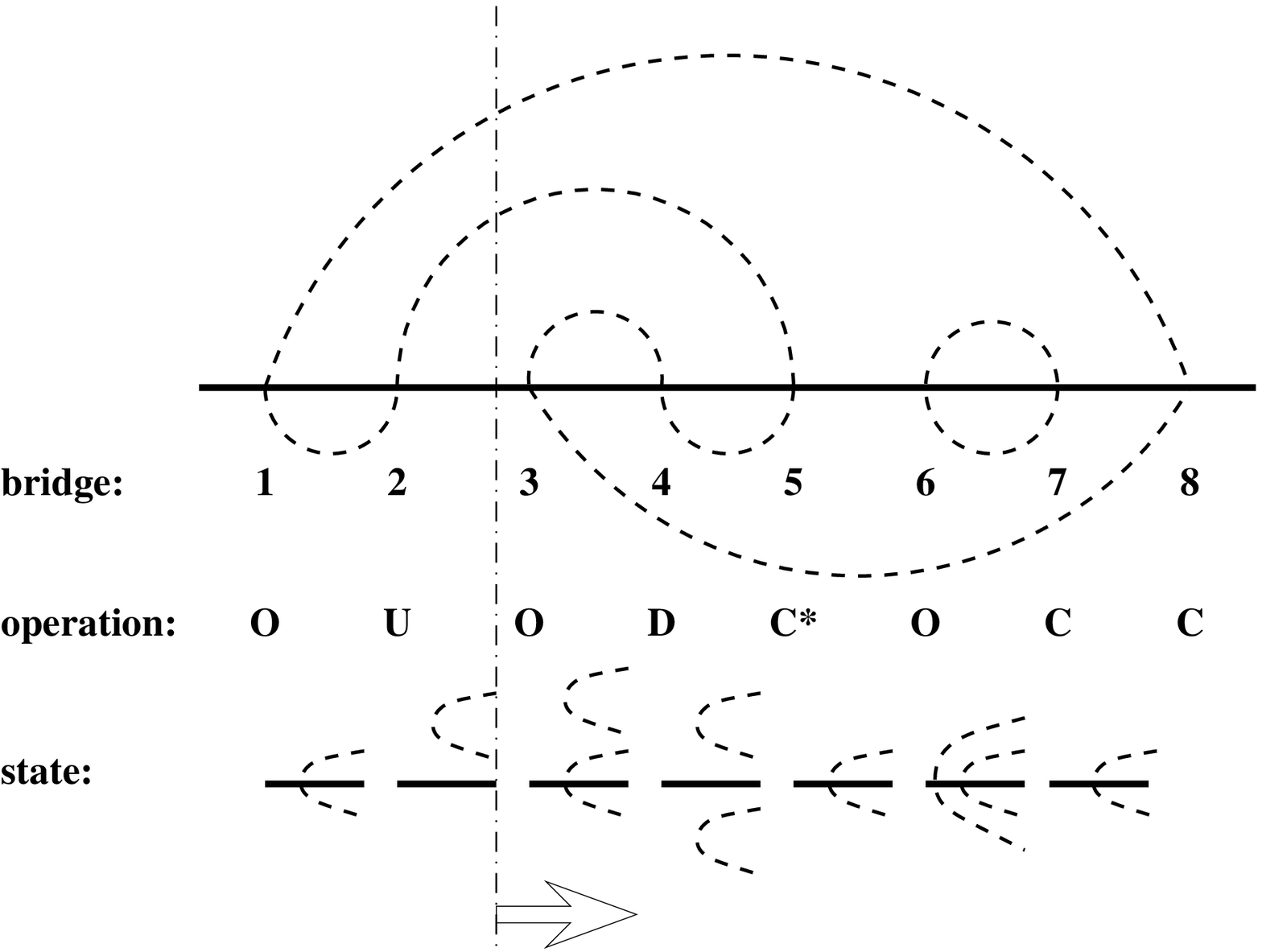}{8.truecm}
\figlabel\fish

Figure~\fish\ shows a typical meander system with one infinite river 
and $N=8$ bridges. The partition function is defined by
giving a weight $n_2=q$ to each road. The transfer matrix acts by transferring
from the left to the right a vertical line intersecting the river between two 
consecutive bridges. A {\em state} characterizes that part of the system
that is to the left of the vertical line by listing the pairwise connectivities
amongst the road segments as well as the position of the river.

Whenever the vertical line is transfered one step to the right, a new bridge 
is added and one out of four possible {\em operations}
(O, C, U and D) can take place, as indicated on the figure. Each of
these operations connects two consecutive states of the system, before and after 
the addition of the bridge. The operation O {\em opens} a new road segment
on top of the river, thus connecting the two sides of the river. Similarly 
the operation C {\em closes} a road on top of the river. This operation comes in two 
variants, depending on whether the closed segment was already connected before 
the addition of the bridge (C) or not (C${}^*$). 
In the former case, the road segment is erased and a non-trivial Boltzmann 
factor of $q$ must be accounted for. In the latter case, the connectivity is 
transformed so as to connect the left-over partners of the two road segments 
that were eliminated. Finally, a road segment immediately below the river can
move {\em up} (operation U), and a segment just above the river can move
{\em down} (operation D).
\medskip
In order to fully specify the transfer matrix of this problem we need
to enumerate the possible connectivity states at a given time.
The number of such states determines the size of the memory needed
to store information at time $t=n$. We will therefore compute  
the number $F(n,N)$ of connectivity states after addition of the $n$-th
bridge for any $n=1,2,\ldots,2N$ for the case of $2N$ bridges.
Also, to implement the calculation of the meander partition function
on a computer, one needs to {\em order} these states so that the
entries of the transfer matrix can be accessed by means of a
one-to-one mapping between the states and the set of integers
$1,2,\ldots,N_{\rm states}$. Here $N_{\rm states}$ stands for the
total number of states encountered in the whole transfer process
from the left of the first bridge to the right of the last one.
It is also the dimension of the explored state space and depends explicitly
on the number $2N$ of bridges added in the whole process.
This number will be estimated below while the explicit ordering procedure
will be presented at the end of this subsection.

\fig{Transfer matrix enumeration of all
 multi-road meanders with 6 bridges. The figure lists the complete set of
 intermediary states after the $n$-th bridge ($n=1,2,\ldots,5$) along with
 their respective weight.}{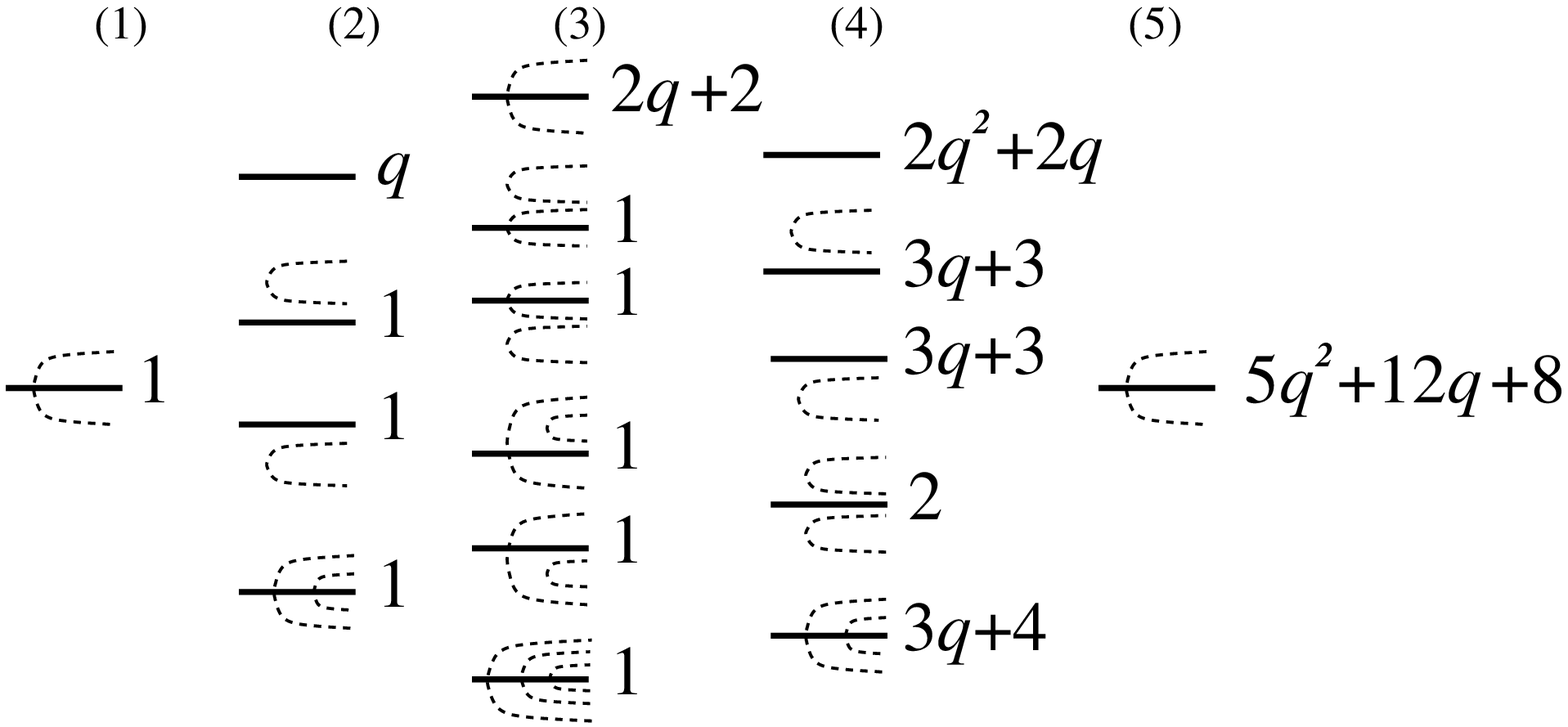}{10.truecm}
\figlabel\states

Before we turn to the general case, let us illustrate the state
counting for one infinite river with $2N=6$ bridges.
In Fig.~\states, we detail the intermediate steps in the
calculation of the corresponding meander polynomial $m_3(q)$. 
The figure depicts
the complete set of states generated by the transfer process, when connecting 
the ``empty" states (vacua) before bridge 1 and after bridge 6, along with 
their respective weights. Apart from the ``empty" states $F(0,6)=F(6,6)=1$, 
we read $F(1,3)=1$, $F(2,3)=4$, $F(3,3)=6$, $F(4,3)=5$, $F(5,3)=1$ and 
$N_{\rm states}=11$. 
The result of the calculation is the meander
polynomial $m_3(q) = 5 q^3 + 12 q^2 + 8 q$, indicating that
with 6 bridges there are respectively 8, 12 and 5 meanders with
1, 2 and 3 connected components.

\noindent{\bf Dimension of the state space:}

Let us now turn to the computation of the numbers $F(n,N)$.
To this end we begin by relating these numbers to
properties of a certain class of restricted Brownian walks.
Consider a situation where there are $p_1$ (resp. $p_2$) road segments above
(resp. below) the river, and where these segments are pairwise connected in 
such a way that exactly $h$ arches cross the river. It is easy to check that 
the four operations O, C, D and U described above always shift $p_1$, $p_2$ and
$h$ by one unit, either $+1$ or $-1$, so that, by induction, 
$n$, $p_1$, $p_2$ and $h$ have the 
same parity. Clearly, we also have $h\le p_1$ and $h\le p_2$ as an arch
crossing the river connects a point above it to a point below it.
There are then $(p_1-h)/2$ (resp. $(p_2-h)/2$) non-crossing arches that 
stay above (resp. below) the river. Now, while a crossing arch can be generated 
by means of a {\it single} bridge, by using the move O, the generation of a 
non-crossing arch necessitates at least {\it two} moves 
(O followed by either U or D). 
Therefore, twice the number of non-crossing arches plus the number
of crossing arches cannot supersede the number of bridges added, i.e. 
$p_1 + p_2 - h \le n$. In terms of $h$, the arch ``height" above the river, 
this leads to the constraint:
\eqn\consth{\max(0,p_1+p_2-n) \le h \le \min(p_1,p_2).} 
Note that the above condition automatically implies that
$p_i\le n$, $i=1,2$.

The above constraint turns out to be the only one as long as $n\le N$.
For $n>N$ there are additional constraints, since we must always be
able to annihilate any given state at level $n$ in at most $2N-n$
moves so as to end up with the ``empty" state at the right of the $2N$-th
bridge . Since $p_1$ and $p_2$ can decrease by at most one at each step,
we thus have to impose for $n>N$ the two extra conditions:
\eqn\extrcons{p_1 \le 2N-n \ ,\ \ \  p_2 \le 2N-n\ .}

\fig{A typical arch configuration (a) below the river and the corresponding
walk (b). The walk has $p=7$ steps and final height $h=3$. 
At each step the height coordinate $h_i$ on (b) changes by $\pm 1$, 
according to whether an arch is opened or closed in (a).}
{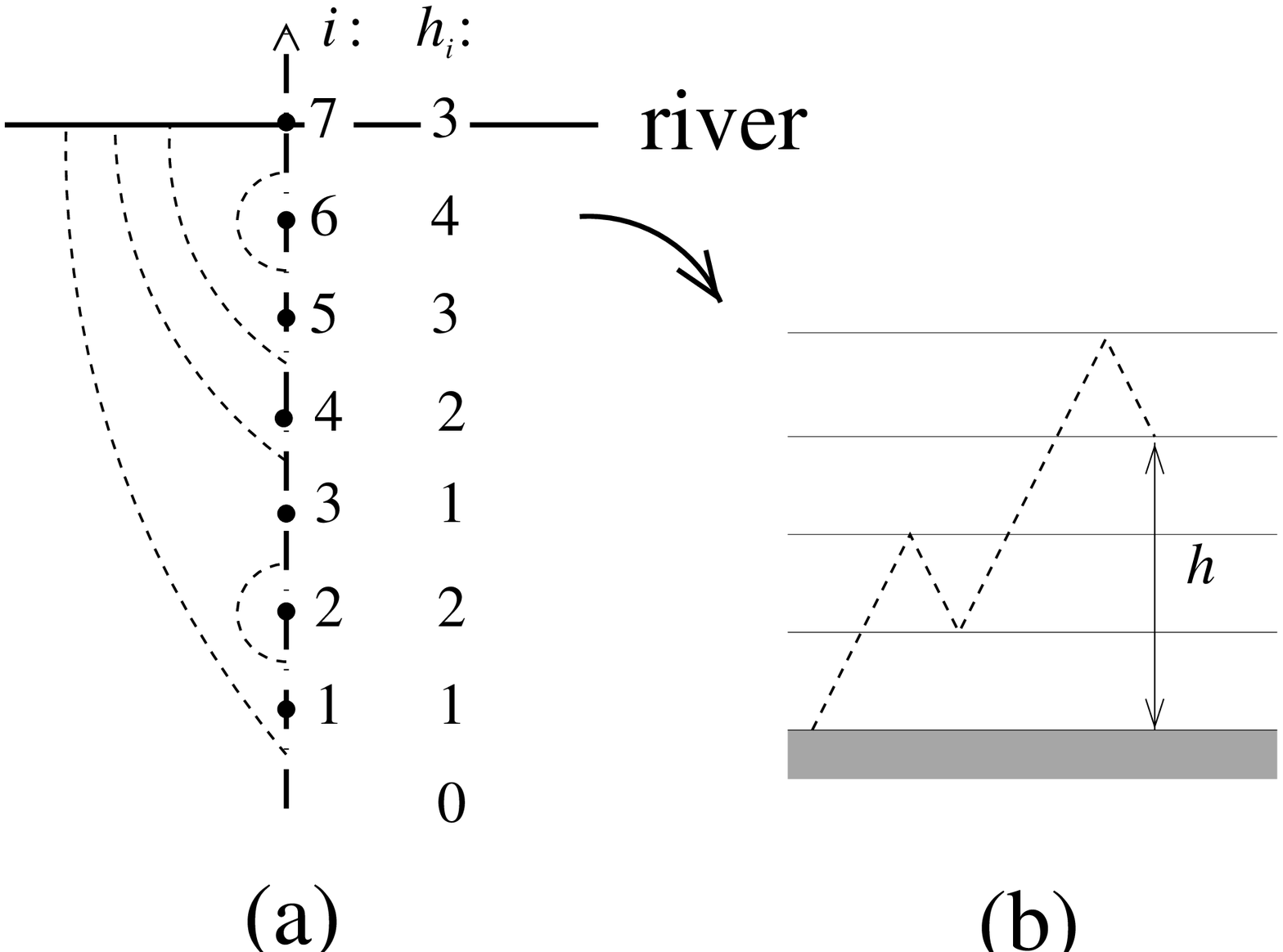}{7.truecm}
\figlabel\walk

It is easy to estimate the number $N_{\rm states}$ of accessible states,
i.e. those which satisfy the above constraints \consth\ and \extrcons\ 
for all $n=1,\ldots,2N$. An arch configuration, read from the bottom and 
upwards, can be mapped onto a $(p_1+p_2)$-step Brownian walk, where the 
position $h_i$ ($i=0,\ldots,p_1+p_2$) of the walk starts from $h_0=0$ and 
is increased (resp. decreased) by one unit for each opening (resp. closing) 
of an arch, as illustrated on Figure~\walk. Evidently these walks are 
constrained by $h_i\ge 0$ for all $i$. Decomposing the walk into its
left part (describing the arches below the river) and its right part
(describing the arches above the river) connected at height $h$, 
we are naturally lead to define $f(p,h)$ as the number of walks going 
from height 0 to height $h$ (or equivalently that of walks going
from height $h$ to height $0$) in $p$ steps.

To evaluate $f(p,h)$ is a standard exercise. First note that without the 
constraint $h \ge 0$ this would simply read ${p \choose (p-h)/2}$ as
me must choose the $(p-h)/2$ ascending steps. Each walk violating the
constraint at some point hits $h_i = -1$. Let $i$ be the first time
this happens, and consider reversing each step from $i$ and
onwards. The result is a walk going from 0 to $(-1)-(h+1)$, or
equivalently from 0 to $h+2$. Therefore:
\eqn\fph{ f(p,h) = {p \choose (p-h)/2} - {p \choose (p-h)/2-1}\ .}

In terms of $f(p,h)$, the number $F(n,N)$ of states accessible at level
$n$ can then be written as
\eqn\FnN{\eqalign{F(n,N) &= \sum_{p_1=0}^{\min(n,N-n)} 
\sum_{p_2=0}^{\min(n,N-n)} \sum_{h=\max(0,p_1+p_2-n)}^{\min(p_1,p_2)}
f(p_1,h) f(p_2,h)
\cr &=\sum_{h=0}^n \sum_{p_1=h}^{\min(n,2N-n)}
          \sum_{p_2=h}^{\min(n+h-p_1,2N-n)} f(p_1,h) f(p_2,h)\ ,}}
where it is implicitly understood that $h$, $p_1$ and $p_2$ in the sums above
all have the parity of $n$.

When $N \ge n$ it is possible to rewrite Eq.~\FnN\ in a simpler
form, since the second constraint in the upper limit of the summations
over $p_1$ and $p_2$ (namely $p_i\le N-n$) does not come into play.
Setting $p_1+p_2=2p$, we simply have to enumerate all the set of $p$ arches, 
cut by the river at a height $h$ that has the same parity as $n$ and 
satisfies the constraint $h \ge 2p-n$. Let us define
\eqn\hmin{h_{\min}(n,p)\equiv \left\{\matrix{\max(0,2p-n) &
\hbox{for $n$ even} \cr \max(1,2p-n) & \hbox{for $n$ odd.}\cr }
\right. }
In the walk language, we have to count 
all the walks of length $2p$ that stay non-negative
with a {\em marking} at a point of height $h \ge h_{\min}$ and with a 
well defined parity, that of $n$.
The total number of non-negative walks of length $2p$ going from $0$ to $0$ 
and with a marked point at position $h$ can be easily calculated to be:
\eqn\gph{g(p,h)=f(2p,2h)+f(2p,2h+2)={2p \choose p-h}-{2p\choose p-h-2}\ .}
Summing over the heights $h\ge h_{\min}(n,p)$ having the parity of $n$
(which is also that of $h_{\min}(n,p)$) we have
\eqn\sumh{\sum_{{h=h_{\min}(n,p)}\atop {h=h_{\min}(n,p) \mod 2}}^{n} g(p,h)=
{2p \choose p-h_{\min}(n,p)}\ .}
The complete number of states used in the transfer matrix 
is now simply obtained by summing over the number of arches.
Depending on the parity of $n$, we get:
\eqn\Feven{ F(n,N \ge n) = \sum_{p=0}^n {2p \choose \min(p,n-p)} \ \ \ \
 \hbox{ for $n$ even,}}
and similarly:
\eqn\Fodd{F(n,N \ge n) = \sum_{p=1}^n {2p \choose \min(p-1,n-p)} \ \ \ \
 \hbox{ for $n$ odd.}}
Note that the above expressions {\it do not depend} on $N$. This is
because we assumed that $n\le N$, in which case the second constraint 
\extrcons\ is ineffective.
For large $n$, the above expressions for $F(n,N\ge n)$ can be evaluated 
by a simple saddle point approximation.
Setting $p=y n$, we get a saddle point at a value $y^\star$ solution 
of 
\eqn\soly{ (3y^\star-1)^3 = (1-y^\star)(2y^\star)^2\ ,}
namely with the numerical value $y^\star=0.611491992\cdots$.
This value indicates that the state statistics is indeed governed by arch
systems with a number $p=p_1=p_2=p^\star\equiv y^\star n$  of arches.
The quantity $F(n,N\ge n)$ is then found to grow
asymptotically as $a^n$, where
\eqn\vala{ a = \frac{(2y^\star)^{2y^\star}}{(1-y^\star)^{1-y^\star} 
(3y^\star-1)^{3y^\star-1}}\ ,}
which, using \soly,  is also the solution of 
\eqn\sola{a^3=(1+a)^2\ .}
This yields the numerical value $a=2.147899036 \cdots$.
\medskip
\vbox{
$$\vbox{\font\bidon=cmr10 \bidon
\offinterlineskip
\halign{\quad \hfill # \tv & \quad \hfill # 
& \hfill # &  \hfill # \ \tv \tv & \quad \hfill # \tv 
& \quad \hfill # & \hfill # & \hfill # \cr
$N$ & $F(N,N)$ & $F(n_{\rm max},N)$ & $n_{\rm max}$ &
$N$ & $F(N,N)$ & $F(n_{\rm max},N)$ & $n_{\rm max}$\cr
\noalign{\hrule}
   1 &           1 &                   1 &             1 &
  13 &       12905 &               26770 &            15 \cr
   2 &           4 &                   4 &             2 &
  14 &       27971 &               62959 &            16 \cr
   3 &           6 &                   6 &             3 &
  15 &       59282 &              155153 &            18 \cr
   4 &          16 &                  16 &             4 &
  16 &      128130 &              388695 &            19 \cr
   5 &          29 &                  29 &           5/6 &
  17 &      272610 &              950128 &            20 \cr
   6 &          68 &                  68 &             6 &
  18 &      588153 &             2279273 &            21 \cr
   7 &         134 &                 161 &             8 &
  19 &     1254586 &             5733997 &            23 \cr
   8 &         300 &                 363 &             9 &
  20 &     2703503 &            14523043 &            24 \cr
   9 &         614 &                 846 &            10 &
  21 &     5777115 &            35946838 &            25 \cr
  10 &        1349 &                1890 &            11 &
  22 &    12438708 &            87192966 &            26 \cr
  11 &        2813 &                4579 &            13 &
  23 &    26613942 &           223196395 &            28 \cr
  12 &        6126 &               11216 &            14 &
  24 &    57268474 &           568622062 &            29 \cr
}}$$
\nobreak
\noindent{\vbox{\baselineskip=12pt \noindent {\bf Table 1:}
 Number of intermediate states
 employed by the transfer matrix algorithm for multi-road meanders with
 $2N$ bridges. The number of states $F(N,N)$ after addition of the
 $N$-th bridge is less than the maximal number of states
 $F(n_{\rm max},N)$, which occurs at a value $n_{\rm max}$ slightly
above $N$.}}
}
\medskip
In Table~1, we give some explicit values of the number of
states needed when enumerating meanders with $2N$ bridges.
It is seen that the maximum number of states occurs slightly after the
addition of the middle bridge, $n=N$.
For large $N$, one can easily estimate the value $n_{\rm max}$ of $n$
where this maximum occurs. Indeed, for $n>N$, the second condition
\extrcons\ starts to play a role. As we have seen, for large $n$,
the state statistics without the second constraint is dominated by arch 
configurations with $p_1=p_2=p^\star=y^\star n$. Therefore, we expect 
this second constraint \extrcons\ to effectively affect the asymptotic 
behavior and start reducing the number of states whenever 
$p^\star=2N-n$, i.e. for $n=2N/(1+y^\star)$. Assuming that this precisely 
corresponds to the step having the maximum number of states, we have 
\eqn\estnmax{n_{\rm max}=2N/(1+y^\star)=1.241085907\cdots N\ .}
We can then estimate
the asymptotic number of encountered states in the whole process 
to grow like:
\eqn\estnstates{N_{\rm states}\sim F(n_{\rm max},N)\sim 
a^{n_{\rm max}}=(2.582603447\cdots)^N\ .}
The estimates \estnmax\ and \estnstates\ agree with the values observed 
in Table 1.
\medskip
\noindent{\bf Ordering the states:}

According to Eq.~\FnN, the full set of $F(n,N)$ states can
be ordered if we know how to order the $f(p,h)$ restricted Brownian
walks defined above. Namely, in that case one can order the complete
set of states lexicographically after $h$, $p_1$, $p_2$, the value of
the first Brownian walk [$1,2,\ldots,f(p_1,h)$], and finally the value
of the second Brownian walk [$1,2,\ldots,f(p_2,h)$]. By this we mean
simply that state $A$ precedes state $B$ if $h_A < h_B$. If $h_A = h_B$
this criterion is inconclusive, and one compares the values of $p_1$,
so that $A$ precedes $B$ if $p_{1,A} < p_{1,B}$. In case of further
equality one proceeds to compare $p_2$, and so on.

To order the Brownian walks, consider first the example contributing to
$f(7,3)$ shown in Fig.~\walk. The idea is to obtain another
formula for $f(p,h)$ which will in turn allow us to define
a recursive ordering of the walks. We start by focusing on the
first (lowermost) arch. Either this arch is open (as is the case on the
figure), or it closes at some other point before $p$. In the first
case the remaining arch configuration is a contribution to
$f(p-1,h-1)$, and in the second the arches inside the first
arch are independent of the arches above its termination
point. We therefore have
\eqn\ffph{f(p,h) = f(p-1,h-1) + \sum_{k=1,3,5,\ldots} f(k-1,0) f(p-k-1,h)\ ,}
with $f(0,h)= \delta_{0,h}$, and $f(p,h)=0$ for $p<0$. This is the required 
formula.

A walk contributing to $f(p,h)$ can now be recursively ordered, first by
considering the termination point of the lowermost arch (which by
definition is infinity if that arch is open), then by (recursively)
considering the ordering of the smaller arch system inside the first
arch, and finally by (recursively) considering the ordering of the
arch system above the first arch (which only exists if that
arch is closed). The procedure just described generalizes the ordering
of the Catalan connectivities $c_{p/2} = f(p,0)$ given in
Ref.~\Blote.

\subsec{Other geometries}

We now come to the case of the more involved meandric geometries 
encountered in Sect.~2.
\medskip
\noindent{\bf Two infinite rivers:}

It is possible to generalize the multi-road one-river transfer matrix
to the case of several rivers, provided that the latter can be deformed
into a system of parallel lines that are only connected among themselves
at infinity. For simplicity we consider in the following the case of
two such infinite rivers. As we have seen, this situation also
corresponds to the deformation of the figure-eight configuration shown 
in Fig.~\eight, provided that a marked point is added on each river.

\fig{The deformation of a two-river configuration by sending all
the bridges crossing the upper river to the left and all the
bridges crossing the lower river to the right. The transfer matrix 
acts by first adding the left bridges, then the right ones.}
{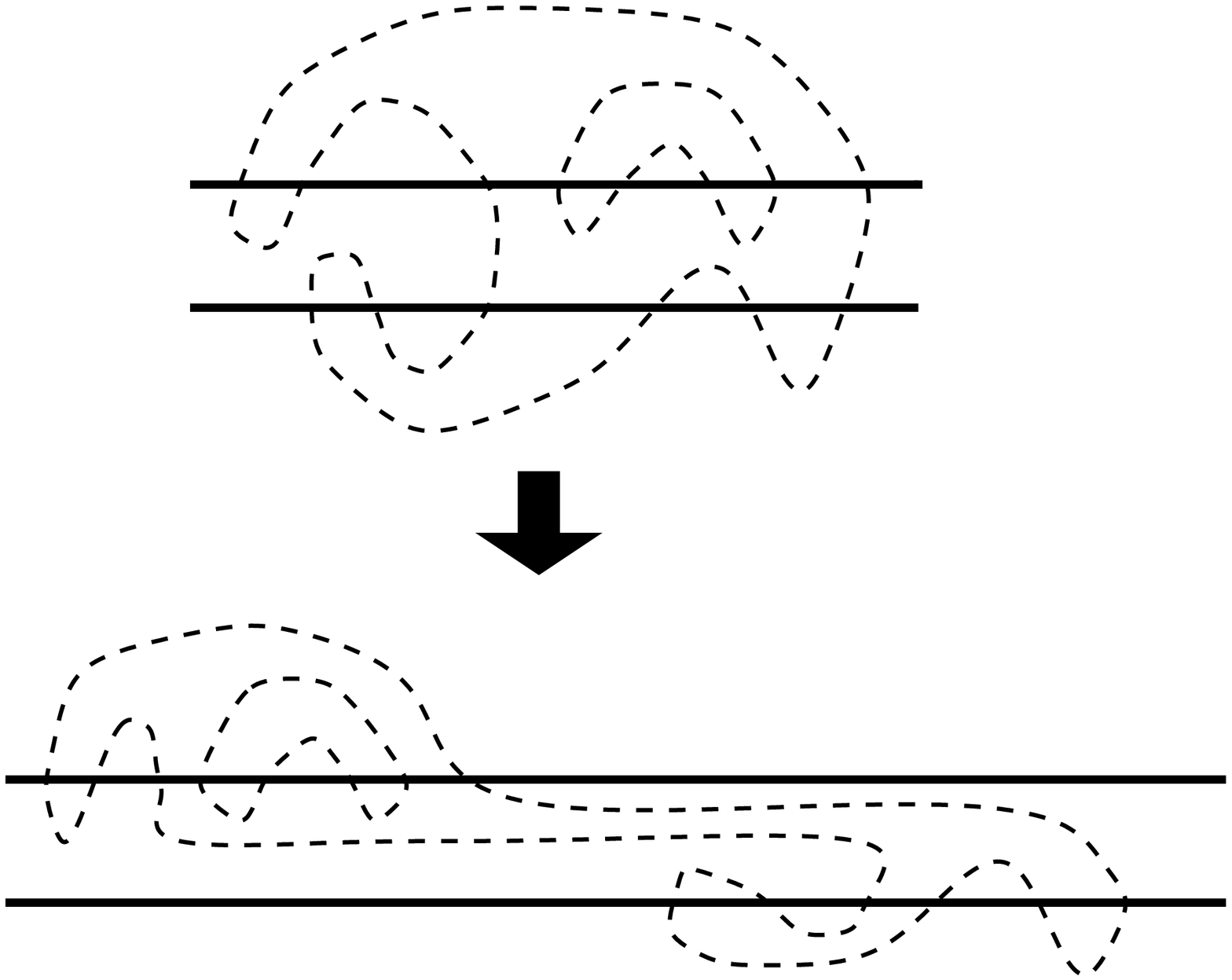}{7.truecm}
\figlabel\tworiv

As before, the crucial point is to have a well-defined transfer direction,
which we simply take to be parallel to both rivers. For any given
configuration with $2N$ bridges we write $N=N_1+N_2$, where $2N_1$
(resp.~$2N_2$) is the number of times the roads cross the first
(resp.~the second) river. Note that contrary to the bridges on the
same river which are naturally ordered, bridges on the first and
on the second river are not naturally ordered with respect to 
one another. To avoid double counting in the transfer matrix approach,
we can simply deform the roads as in Fig.~\tworiv\ so as to send all 
the bridges of river $1$
to the left and all those of river $2$ to the right. In the transfer process,
we thus add first the $2N_1$ bridges crossing the first river and then
the $2N_2$ bridges crossing the second river.

Given a decomposition $N=N_1+N_2$,
the calculations therefore proceed exactly as in the one-river case,
except that when both $N_1>0$ and $N_2>0$ not {\em all} of the
connectivity states described in Sect.~3.1 come into use.
Denoting by $p_1$ the number of road segments above the first river,
by $p_2$ the road segments in between rivers, and by $p_3$ the segments
below the second river, we now have the following constraints after the
addition of the $n$-th bridge:
\eqn\pis{\eqalign{
  p_1 &\le \max \left( \min(n,2N_1-n),0 \right),  \cr
  p_2 &\le \min(n,2N-n), \cr
  p_3 &\le \max \left( \min(n-2N_1,2N-n),0 \right)\ .\cr}
}

The total two-river meander polynomial is obtained by summing over the
possible decompositions. We can furthermore simply address the situation with
a marked point on either river by weighing each
term in the decomposition by a factor of $(2N_1+1)(2N_2+1)$.
\medskip
\noindent{\bf One semi-infinite river (semi-meanders):}

Finally, we have examined systems of semi-meanders, where the roads are
allowed to wind around the source of a semi-infinite river. 
Each winding number $w$
can be examined separately, by choosing an initial state of $w$ arches
nested inside one another, and simply applying the four operations discussed
in Sect.~3.1. Unfortunately in this situation we have not found a
simple and full description of the generated states 
(apart from $w=0$ and $w=1$).
It is however still possible to carry out the transfer matrix
calculations, by simply inserting the generated states in an unordered
list. The price to be paid is that to find a given state one will have
to sequentially search through the entire list, leading to a pitiful waste
of computation time. This is nevertheless what we have done, and 
consequentially we have had to content ourselves with smaller system
sizes.

\subsec{Tangency points}

\fig{The six new operations allowed for tangent meanders.}{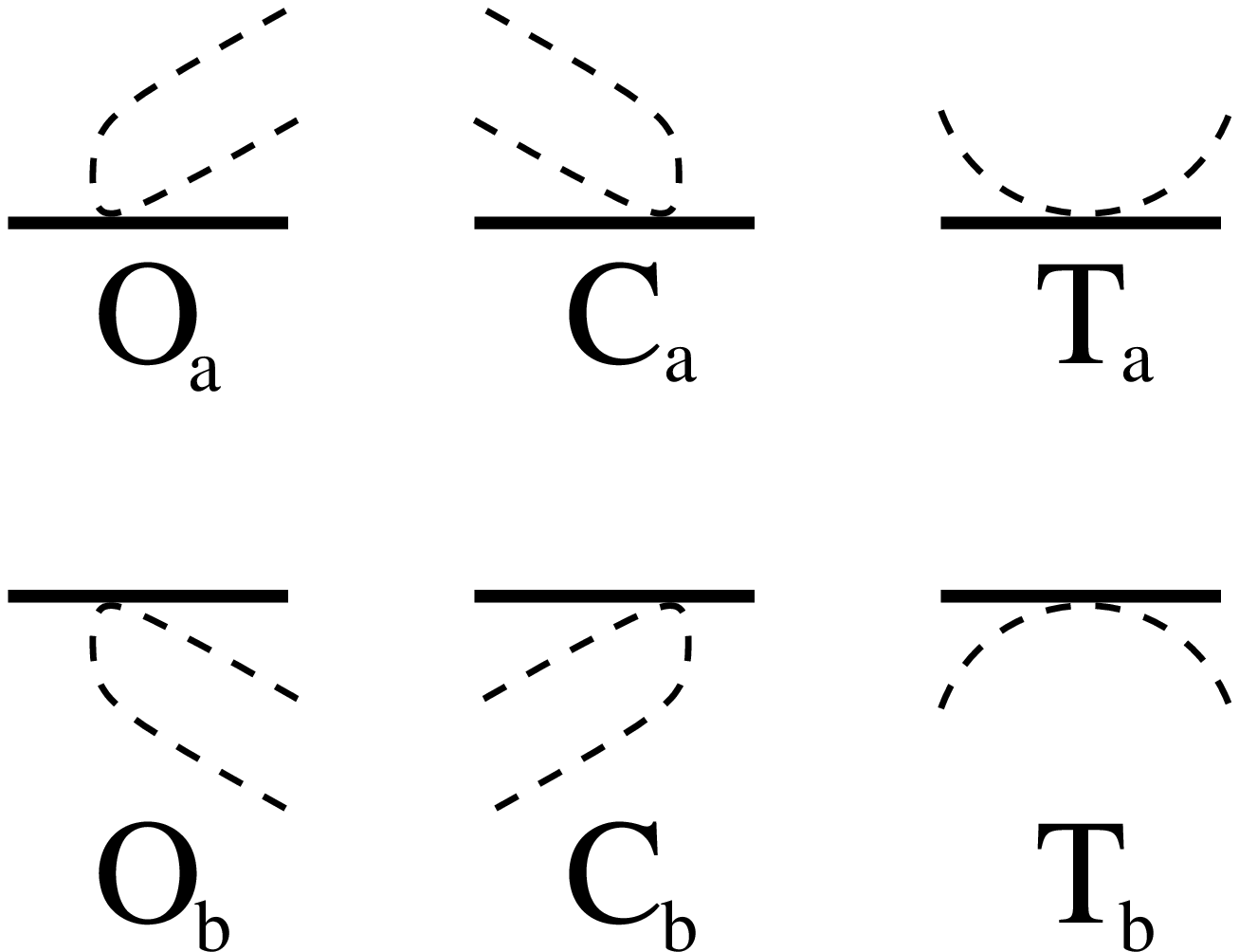}{5.truecm}
\figlabel\tangence

We have also enumerated systems of {\em tangent} meanders, where apart
from the four types of moves at each bridge described in
subsection 3.1, we allow for the six additional operations depicted
in Figure~\tangence.
These moves allow to {\em open} a road whilst staying {\em above} or
{\em below} the river (${\rm O}_{\rm a}$ and ${\rm O}_{\rm b}$),
to similarly {\em close} a road without crossing the river
(${\rm C}_{\rm a}$ and ${\rm C}_{\rm b}$), and finally to {\em touch}
the river whilst staying {\em above} it or {\em below} it
(${\rm T}_{\rm a}$ and ${\rm T}_{\rm b}$).
The meander polynomial is now defined by assigning a weight $x$ to each
of the crossing bridges and a weight $y$ to each of the tangent vertices.

Here again, we have not been able to find an explicit way of enumerating
and ordering the states obtained by applying this set of ten operations to
the vacuum. 

\newsec{Numerical results}

Let us now present our numerical results obtained by use of the algorithms
presented in Sect.~3, together with an appropriate extrapolation 
procedure. Subsection 4.1 discusses multi-component meanders and introduces
the extrapolation method. Subsection 4.2 is devoted to the case of two
parallel rivers, while subsection 4.3 addresses semi-meanders. Finally,
we study tangent meanders in subsection 4.4. 

\subsec{One infinite river (meanders)}

We have enumerated multi-connected meanders with one infinite river 
and a {\it fixed fugacity} $q$ up to $N=24$, i.e. $2N=48$ bridges for
$q=0$, $\sqrt{2}$, $\sqrt{3}$ and $2$. In other words, we evaluated
the quantities $M_N=\lim_{q\to 0}(m_N(q)/q)$, $m_N(\sqrt{2})$, 
$m_N(\sqrt{3})$ and $m_N(2)$ for $N=1,\ldots,24$.
{}From these numbers, we can extract estimates for the large $N$ 
``activity" per bridge $R(q)$ and configuration exponent 
$\alpha(q)$ defined as in Eq.~\larbeha\ by :
\eqn\larbehabis{m_N(q)\sim C(q) {R(q)^{2N}\over N^{\alpha(q)}}\ .}
The estimates for $\alpha(q)$ can be transformed into estimates for the 
central charge $c(q)$ through the following relation, inverting \kpz\
for $\gamma=2-\alpha(q)$:
\eqn\cdeq{c(q)=1-6{(2-\alpha(q))^2\over (\alpha(q)-1)}\ .}
Starting from the zero-th order values:
\eqn\zerothest{\eqalign{R_N^{(0)}(q)&\equiv\sqrt{m_{N+1}(q)\over m_N(q)}\cr
\alpha_N^{(0)}(q)&\equiv (N+1)^2\left({m_{N+2}(q)m_N(q)\over (m_{N+1}(q))^2}-1
\right)\ ,\cr}}
we can build better estimates $R_N^{(p)}$ and $\alpha_N^{(p)}$ by
a recursive use of a standard convergence acceleration procedure:
\eqn\nevil{R^{(p)}_N=R^{(p-1)}_N-{p+1\over p}
\ {(R^{(p-1)}_{N+1}-R^{(p-1)}_N)
(R^{(p-1)}_N-R^{(p-1)}_{N-1}) \over R^{(p-1)}_{N+1}+R^{(p-1)}_{N-1}
-2 R^{(p-1)}_N}}
and similarly for $\alpha_N^{(p)}$.
The procedure implicitly
assumes that the corrections to the asymptotic scaling \larbehabis\ are
regular, i.e. integer powers of $1/N$, in which case it guarantees that 
$R_N^{(p)}(q)=R(q)+{\cal O}(1/N^{p+1})$.

\fig{The estimated central charge obtained from the values 
$\alpha_N^{(p)}(q)$ with $p=1,2,3$ and $4$ iterations. The
size of the symbols decreases with the number of iterations. Each
value is represented at an abscissa $n$ corresponding to the
largest index $N=n$ of $m_N(q)$ used in the estimate. We also draw horizontal 
lines at the predicted values of $c(q)$.}{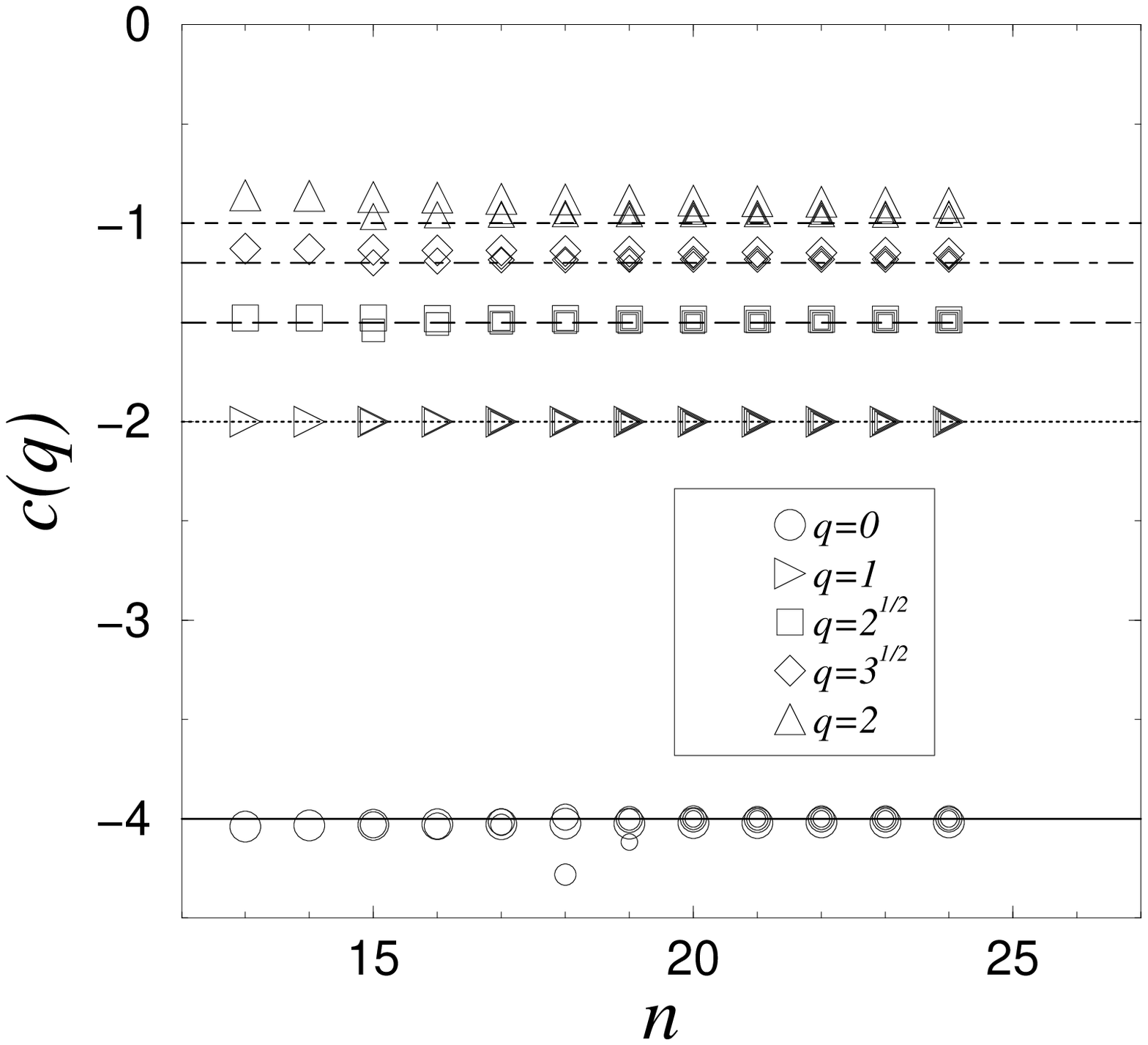}{9.truecm}
\figlabel\resoneriv

Fig.~\resoneriv\ shows the estimated central charge obtained through
\cdeq\ using $\alpha^{(p)}_N(q)$ for $p=1,2,3,4$, together
with the expected central charge according to \newchar\ with $n_1=0$
and $n_2=q$. For completeness, we also show the corresponding estimates 
obtained for $q=1$ with $m_N(1)=(c_N)^2$ in terms of the Catalan numbers 
$c_N$.  The corresponding quantitative values are displayed in Table 2 below.
\medskip
\vbox{
$$\vbox{\font\bidon=cmr10 \bidon
\offinterlineskip
\halign{\quad \hfill # \tv \tv & \hfill # \hfill \tv
& \quad \hfill # \hfill \tv \tv & \quad \hfill # \hfill \tv \tv & 
\quad \hfill # \hfill  \tv & \quad \hfill # \hfill\tv \tv \cr
$q$ & $\alpha_{\rm theor.}(q)$ & $\alpha_{\rm num.}(q)$ & $R^2_{\rm num.}(q)$ &
$c_{\rm theor.}(q)$ & $c_{\rm num.}(q)$ \cr
\noalign{\hrule}
0 & ${29+\sqrt{145}\over 12}=3.4201328\cdots$ & 3.4207 & 12.26286 & -4 & -4.003 \cr
1 & 3 & 3.00000 & 16.000000 & -2 & -2.0000 \cr
$\sqrt{2}$ & ${53+\sqrt{265}\over 24}=2.8866175\cdots$ & 2.885 & 17.52468 & -3/2 & -1.496 \cr
$\sqrt{3}$ & ${131+\sqrt{1441}\over 60}=2.8160084\cdots$ & 2.812 & 18.68970 & -6/5 &  -1.18 \cr
2 & ${13+\sqrt{13}\over 6}=2.7675918\cdots$ & 2.75 & 19.669 & -1 & -0.95 \cr
}}$$
\nobreak
\noindent{\vbox{\baselineskip=12pt \noindent {\bf Table 2:}
 Numerical estimates for the meander configuration exponent $\alpha(q)$,
the activity $R^2(q)$ per pair of bridges, and the central charge $c(q)$.
The error is implicitly on the last digit. 
The corresponding theoretical values are also listed.}}}
\medskip
\fig{Numerical estimates (triangles) for the meander configuration exponent 
$\alpha(q)$ for $0\leq q \leq 2$, together with the predicted
value \almeq\--\charge\ (solid line). 
We display the result of several iterations
of our convergence acceleration procedure. 
The size of the symbols decreases with 
the number of iterations.}{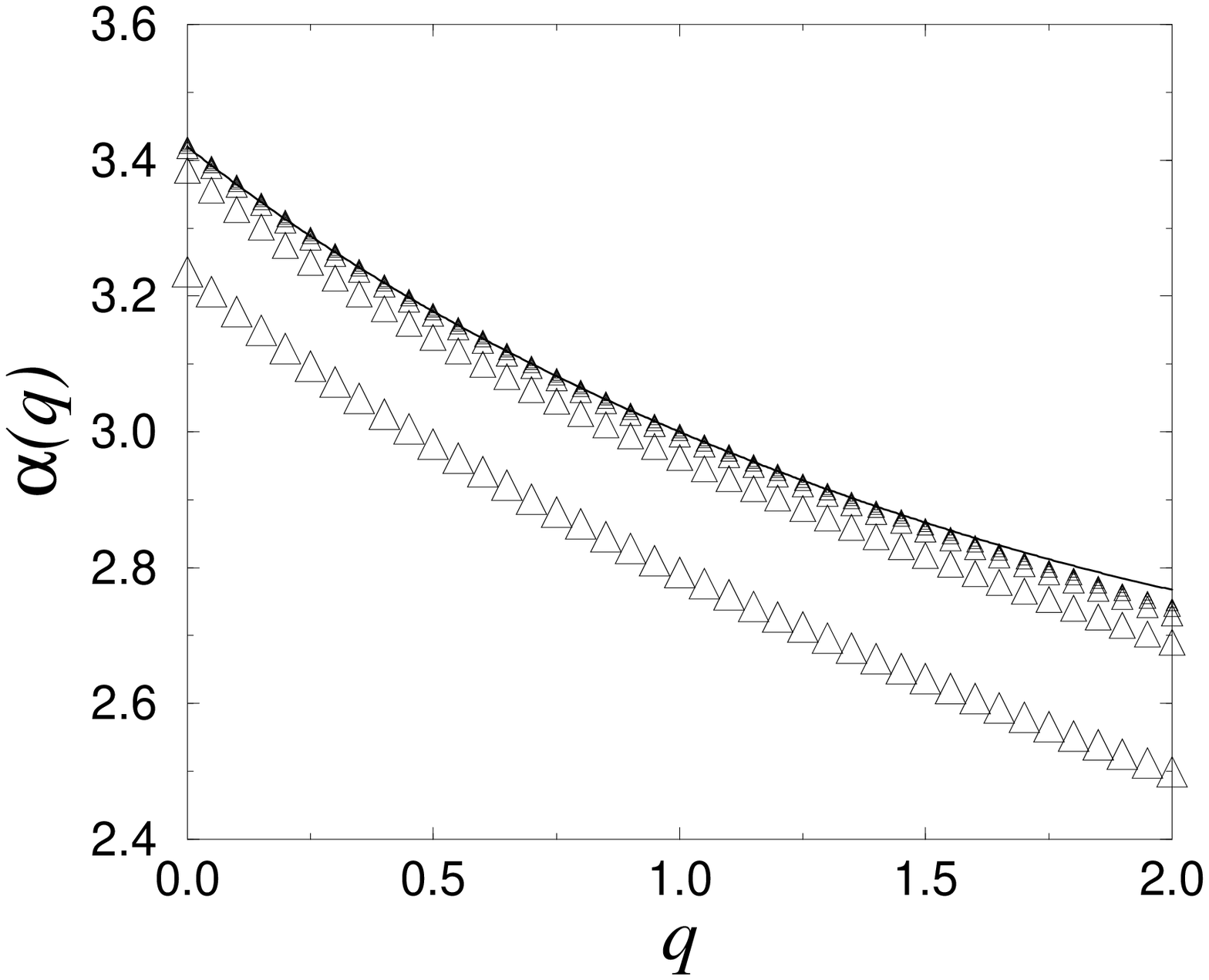}{9.truecm}
\figlabel\alphadat

We have also computed the numbers $M_N^{(k)}=\mu_{2N,0}^{(k)}$ of meanders
with one infinite river, $2N$ bridges and 
a fixed number $k$ of connected components ($k=1,\ldots,N$)
for $N=1,\ldots,20$ ($40$ bridges). 
As an illustration we display the results for $N=20$ in Table 3 below.

\medskip
\vbox{
$$\vbox{\font\bidon=cmr10 \bidon
\offinterlineskip
\halign{\quad \hfill # \tv & \quad \hfill # \tv\tv
& \hfill # \tv &  \hfill # \cr
$k$ & $M_{20}^{(k)}$ &$\ \ $ $k$ & $M_{20}^{(k)}$\cr
\noalign{\hrule}
  1 &     64477712119584604   &          11 &  $\ \ $  706958959835806990         \cr
  2 &     511563350415103008   &               12 &   235265604762448572          \cr
  3 &     1901345329566422790   &               13 &  64713641205591820           \cr
  4 &     4405839231880790648   &               14 &  14658557362753320           \cr
  5 &     7145814923879522986   &               15 &  2709804590263296           \cr
  6 &     8632733743310196256   &               16 &  402058856155712           \cr
  7 &     8070705247685170684   &               17 &  46500885666900           \cr
  8 &     5988061883039308848   &               18 &  3978168316200           \cr
  9 &     3587066097601934530   &               19 &  226760523600           \cr
 10 &     1755310029771295216   &               20 &  6564120420           \cr
}}$$
\nobreak
\noindent{\vbox{\baselineskip=12pt \noindent {\bf Table 3:}
Multi-component meander numbers $M_N^{(k)}$ with 
$2N=40$ bridges and $k$ connected components
of road, $k=1,2,\ldots,20$.}}
}
\medskip

From these values, we can extract 
the polynomials
$m_N(q)=\sum_{k=1}^N M_N^{(k)}q^k$, hence the values $R(q)$ and 
$\alpha(q)$ for a {\it varying fugacity} $q$. These values are displayed
in Fig.~\alphadat\ together with the prediction \almeq\--\charge.

All our estimates clearly validate the theoretical predictions
with a very good accuracy. We note a small discrepancy for values of 
$q$ close to $q=2$. This corresponds to the regime which has the 
worst convergence of our acceleration procedure. This poor convergence
might be due to either corrections which are not regular, as 
implicitly assumed, or to the vicinity of a transition at $q=2$ 
where the DPL$^2$ model stops being critical.
We anyway impute this small discrepancy to our estimation procedure. 

\subsec{Two infinite rivers}

Beyond the central charge, we can also test the operator 
content of the theory. A first check concerns the dimension 
$\Delta_4$ of the operator creating a vertex with four outgoing
river segments, which, as we already explained, can be measured
through the configuration exponent $\alpha_{\rm 2-mark}$ 
for two parallel infinite rivers with a marked point on 
each river. As we mentioned before, in the absence of marking,
we expect the two-river configuration exponent $\alpha_{\rm 2-riv.}(q)$
to be simply the same as that of meanders with a single infinite
river.

\fig{The estimated configuration exponent for two {\it un-marked}
rivers,  obtained with $p=1,2,3$ and $4$ iterations of the 
acceleration procedure \nevil . 
The size of the symbols decreases with the number of iterations. Each
value is represented at an abscissa $n$ corresponding to the
largest index $N=n$ of $m_N^{\rm 2-riv.}(q)$ 
used in the estimate. We also draw horizontal 
lines at the predicted values of 
$\alpha_{\rm 2-riv.}(q)$.}{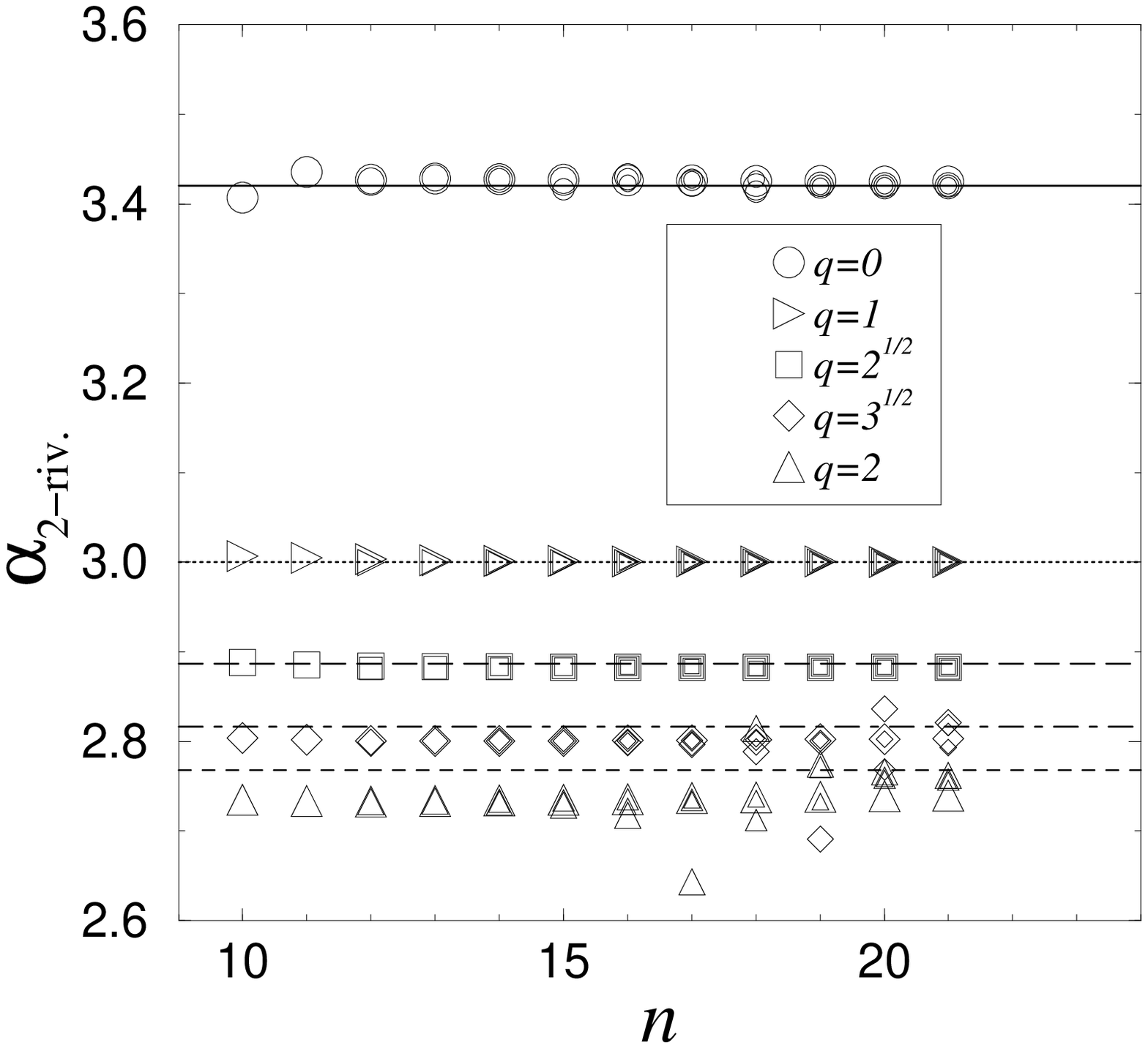}{9.truecm}
\figlabel\restworiv

\fig{The estimated configuration exponent for two {\it marked}
rivers,  obtained with $p=2,3$ and $4$ iterations of the (modified,
as explained in the text) acceleration procedure. 
The size of the symbols decreases with the number of iterations. Each
value is represented at an abscissa $n$ corresponding to the
largest index $N=n$ of $m_N^{\rm 2-mark}(q)$ 
used in the estimate. We also draw horizontal 
lines at the predicted values of 
$\alpha_{\rm 2-mark}(q)$.}{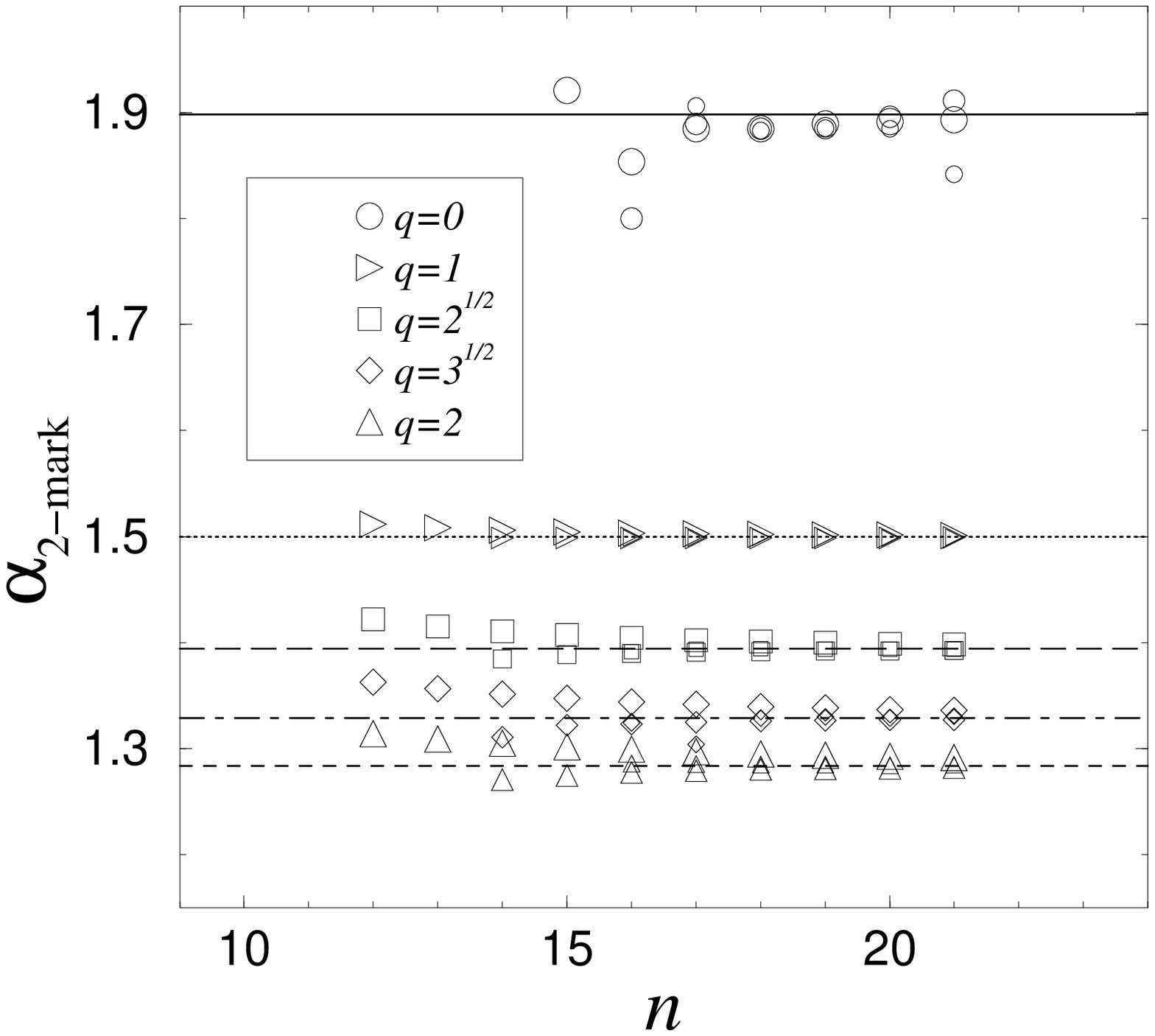}{9.truecm}
\figlabel\restwomark

Fig.~\restworiv\ and \restwomark\ show our results based on 
enumerations with up to $2N=42$ bridges for fixed $q=0$, $\sqrt{2}$,
$\sqrt{3}$ and $2$. For completeness, we also added estimates 
for $q=1$, as extracted from the exact expressions:
\eqn\mntworiv{m_N^{\rm 2-riv.}(q=1)=\sum_{p=0}^{N}c_p c_{N-p} c_N=c_Nc_{N+1}}
for two un-marked rivers, and
\eqn\mntwomark{m_N^{\rm 2-mark}(q=1)=\sum_{p=0}^{N}(2p+1)(2(N-p)+1)c_p 
c_{N-p} c_N =c_N\left(4^{N+1}-{2n+3\choose n+1} \right)}
for two marked rivers. Note that in this latter case of two
marked rivers, the leading contribution 
$m_N^{\rm 2-mark}(1)\sim (16)^N/N^{3/2}$ is corrected by
a sub-leading contribution $\sim (16)^N/N^2$, with an exponent displaced
by a half-integer. In this case, the corrections to the leading
term thus involve generically half-integers instead of integers. We expect 
this phenomenon to be generic for all values of $q$ with a double expansion
involving both the exponent $\alpha_{\rm 2-mark}(q)$ and its ``descendants"
shifted by integers, and a sub-leading correction with the exponent 
$\alpha_{\rm 2 riv.}(q)-1$ and its descendants. Since $\alpha_{\rm 2-riv.}
-1-\alpha_{\rm 2-mark}$ is always close to $1/2$ for the values of $q$
of interest, we decided to use a modified acceleration procedure assuming 
half-integer corrections to scaling (the index $p$ in \nevil\ now takes
half-integer values). The estimates of Table 4 and Fig.~\restwomark\ 
have been obtained with this modified procedure.
\medskip
\vbox{
$$\vbox{\font\bidon=cmr10 \bidon
\offinterlineskip
\halign{\quad \hfill # \tv \tv & \hfill # \hfill \tv
& \quad \hfill # \hfill \tv \tv & \quad \hfill # \hfill \tv \tv \cr
$q$ & $\alpha_{\rm 2-riv.}(q)$ & $\alpha_{\rm num.}(q)$ & $R^2_{\rm num.}(q)$\cr
\noalign{\hrule}
0 & ${29+\sqrt{145}\over 12}=3.4201328\cdots$ & 3.4205 & 12.2627 \cr
1 & 3 & 3.00000 & 16.000000 \cr
$\sqrt{2}$ & ${53+\sqrt{265}\over 24}=2.8866175\cdots$ & 2.882 & 17.5246 \cr
$\sqrt{3}$ & ${131+\sqrt{1441}\over 60}=2.8160084\cdots$ & 2.80 & 18.688 \cr
2 & ${13+\sqrt{13}\over 6}=2.7675918\cdots$ & 2.75 & 19.663 \cr
\noalign{\hrule}
\noalign{\hrule}
$q$ & $\alpha_{\rm 2-mark}(q)$ & $\alpha_{\rm num.}(q)$ & $R^2_{\rm num.}(q)$\cr
\noalign{\hrule}
0 & ${\sqrt{5}+\sqrt{14}\over \sqrt{29}-\sqrt{5}}=1.8982348\cdots$ & 1.89 & 12.26 \cr
1 & 3/2 & 1.499 & 15.9999 \cr
$\sqrt{2}$ & ${\sqrt{5}+\sqrt{23}\over \sqrt{53}-\sqrt{5}}=1.3941001\cdots$ & 1.393 & 17.527 \cr
$\sqrt{3}$ & ${\sqrt{11}+\sqrt{56}\over \sqrt{131}-\sqrt{11}}=1.3285858\cdots$ & 1.33 & 18.69 \cr
2 & ${\sqrt{11}+\sqrt{2}\over \sqrt{26}-\sqrt{2}}=1.2838772\cdots$ & 1.28 & 19.67 \cr
}}$$
\nobreak
\noindent{\vbox{\baselineskip=12pt \noindent {\bf Table 4:}
 Numerical estimates for the configuration exponent and
the activity $R^2(q)$ per pair of bridges in the case of
two un-marked rivers (upper half) and two marked rivers (lower
half).  The error is implicitly on the last digit. 
The corresponding theoretical values are also listed.}}}
\medskip
Here again, the numerical values corroborate our predictions
for the various configuration exponents. Note that we find that
the activity per bridge is for these two cases the same as 
for a single infinite river. It should not come as a surprise
that the activity $R(q)$ per bridge is independent of the particular 
meander geometry since it is a bulk quantity insensitive to 
the choice of boundary conditions (river shapes). It is also
interpreted as the inverse of the convergence radius common 
to all the generating functions of the numbers at hand, also viewed
as various correlators within the same theory. 

\subsec{One semi-infinite river (semi-meanders)}

\fig{Numerical estimates for the semi-meander configuration exponent 
${\bar \alpha}(q)$ for $0\leq q \leq q_{\rm c}$, together with the predicted
value \alsm\--\charge. We display the result of several iterations
of the acceleration procedure. The size of the symbols decreases with 
the number of iterations.}{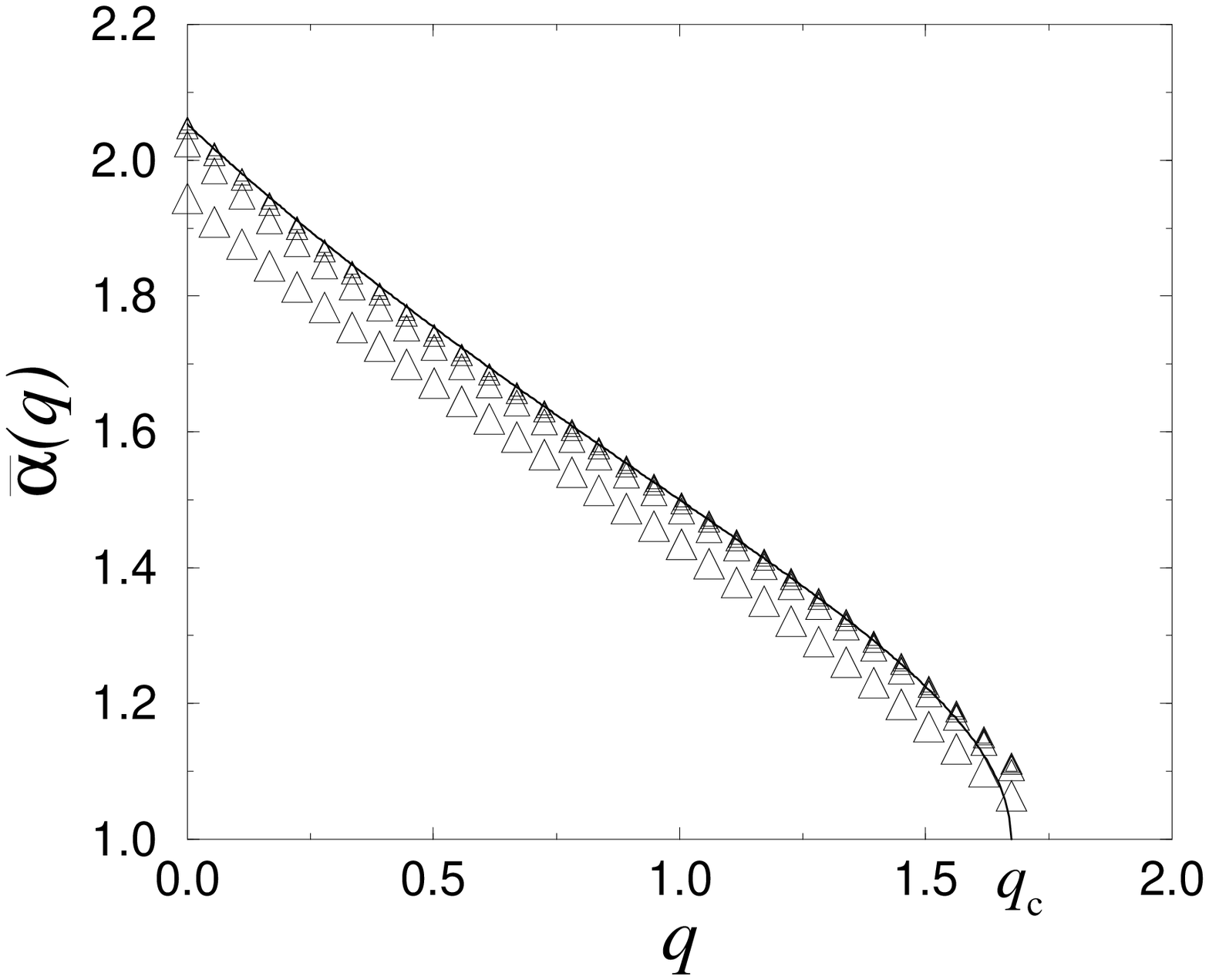}{9.truecm}
\figlabel\alphabardat

As a second check of the operator dimensions of the theory, we 
also considered the case of semi-meanders with a single semi-infinite
river ($n_1=0$) and a arbitrary number $k$ of  connected components of road.
Our estimates rely on an exact enumeration of the numbers ${\bar M}_N^{(k)}(w)$
of semi-meanders with $N$ bridges, $k$ roads and winding number $w$ 
for $1\leq N\leq 33$, $1\leq k\leq N$, $0\leq w \leq N$ and $w=N$ mod 2. 
For illustration, we list in Table 5 below respectively the numbers 
${\bar M}_{33}^{(k)}=\sum_w {\bar M}_{33}^{(k)}(w)$ of semi-meanders with $N=33$ bridges
and $k$ roads (irrespectively of the winding numbers) as well as the numbers
${\bar M}_{33}^{(1)}(w)$ of connected semi-meanders with winding 
number $w=1,3,5,\ldots,33$.

\medskip
\vbox{
$$\vbox{\font\bidon=cmr10 \bidon
\offinterlineskip
\halign{\quad \hfill # \tv & \quad \hfill # \tv\tv
& \hfill # \tv &  \hfill # \cr
$k$ & ${\bar M}_{33}^{(k)}$ &$\ \ $ $k$ & ${\bar M}_{33}^{(k)}$\cr
\noalign{\hrule}
  1 &   455792943581400     &          18 & $\ \ $ 3659252585228            \cr
  2 &   3285874327160852     &          19 &       1040041120124      \cr
  3 &   11119764476127424     &          20 &      279039302088       \cr
  4 &   23598381333433844     &          21 &      70513532334       \cr
  5 &   35436190513634790     &          22 &      16729859124       \cr
  6 &   40334792072264540     &          23 &      3710923316       \cr
  7 &   36464182713722576     &          24 &      765325248       \cr
  8 &   27136413723456560     &          25 &      145710912       \cr
  9 &   17127401092409102     &          26 &      25374900       \cr
  10 &  9409371247346540      &          27 &      3992846       \cr
  11 &  4602479751584184      &          28 &      558396       \cr
  12 &  2042918178657320      &          29 &      67804       \cr
  13 &  835326688674886      &          30 &       6904      \cr
  14 &  318096906554664      &          31 &       557      \cr
  15 &  113643690324368      &          32 &       32      \cr
  16 &  38261586556480      &          33 &        1     \cr
  17 &  12168938393766      &           &             \cr
\noalign{\hrule}
\noalign{\hrule}
$w$ & ${\bar M}_{33}^{(1)}(w)$ &$\ \ $ $w$ & ${\bar M}_{33}^{(1)}(w)$\cr
\noalign{\hrule}
  1 &   59923200729046     &          19 & $\ \ $ 16277801502             \cr
  3 &   121544501379440     &          21 &      1326698396        \cr
  5 &   125267070807626     &          23 &      73827420        \cr
  7 &   85716694410306     &          25 &       2638462       \cr
  9 &   42336073574012     &          27 &       55052       \cr
  11 &   15599486790514     &          29 &      568        \cr
  13 &   4337132101822     &          31 &         2     \cr
  15 &   908960663970     &          33 &        0      \cr
  17 &   142142103262     &           &              \cr
}}$$
\nobreak
\noindent{\vbox{\baselineskip=12pt \noindent {\bf Table 5:}
Multi-component semi-meander numbers ${\bar M}_N^{(k)}$ with
$N=33$ bridges and $k$ connected components
of road, $k=1,2,\ldots ,33$, and fixed winding connected semi-meander numbers
${\bar M}_N^{(1)}(w)$ with $N=33$ and $w=1,3,5,\ldots ,33$.}}
}
\medskip

For this particular geometry,
a clear parity effect occurs and it is thus important to make numerical
evaluations from semi-meander numbers with either even or odd $N$ to avoid large
numerical errors. The results presented in Fig.~\alphabardat\ 
were extracted from
even values of $N$. Table 6 lists quantitative values for the configuration 
exponent and the activity per pair of bridges. Here again, we note
that the activity per bridge is identical to that of the other geometries. 
As shown in Fig.~\alphabardat , the configuration
exponent is in very good agreement for $q$ less than $~1.5$. For
larger values of $q$, a small discrepancy appears, again imputable
to a poor convergence probably due to the
vicinity of a transition point, now at $q=q_{\rm c}$ where we expect a 
proliferation of connected components of road for the semi-meander 
geometry. 
\medskip
\vbox{
$$\vbox{\font\bidon=cmr10 \bidon
\offinterlineskip
\halign{\quad \hfill # \tv \tv & \hfill # \hfill \tv
& \quad \hfill # \hfill \tv & \quad \hfill # \hfill \tv \tv & 
\quad \hfill # \hfill \tv & \quad \hfill # \hfill \tv \tv \cr
$q$ & ${\bar \alpha}_{\rm theor.}(q)$ & ${\bar \alpha}_{\rm even}(q)$ 
& ${\bar \alpha}_{\rm odd}(q)$ & ${\bar R}^2_{\rm even}(q)$ & 
${\bar R}^2_{\rm odd}(q)$ \cr
\noalign{\hrule}
0 & ${1+{\sqrt{11}\over 24}(\sqrt{29}+\sqrt{5})\atop \ \ \ =2.0531987\cdots}$ & 2.0532 & 2.051 & 12.26287 & 12.2626 \cr
1 & 3/2 & 1.500000 & 1.500000 & 16.000000 & 16.000000 \cr
$\sqrt{2}$ & ${1+{\sqrt{2}\over 48}(\sqrt{53}+\sqrt{5})\atop \ \ \ =1.2803730\cdots}$
 & 1.282 & 1.282 & 17.5247 & 17.5247 \cr
}}$$
\nobreak
\noindent{\vbox{\baselineskip=12pt \noindent {\bf Table 6:}
Numerical estimates for the configuration exponent and
the activity $R^2(q)$ per pair of bridges in the case of
one semi-infinite river (semi-meanders).  The error is 
implicitly on the last digit. The corresponding theoretical values are 
also listed.}}}
\medskip
\subsec{Tangent meanders}

Finally, we also checked our assertion that the ``tangency" vertices 
of type b (see Fig.~\twovert ) are irrelevant. 
We computed the canonical partition function
$\mu_N(q;x,y)$ defined in \pfmu\ for the particular case of tangent
meanders with a single connected component of road ($q=0$), a fixed 
weight $x=1$ for ``crossing" vertices and with several values of the 
weight $y$ per ``tangency" vertex. We enumerated up to $N=24$ 
vertices for $y=0.5$, $1$ and $2$ while $y=0$ simply corresponds to 
the meander configurations of subsection 4.1 for which we have results 
up to $48$ vertices. 

\fig{The estimated configuration exponent $\alpha$ for tangent
meanders with a weight $y=0$, $0.5$, $1$ and $2$ per tangency vertex.
The size of the symbols decreases with the number $p$ of iterations
(here $p=1,2,3,4$). Each value is represented at an abscissa 
$n$ corresponding to the largest index $N=n$ for the number of vertices 
(resp. the number of bridge pairs in the case $y=0$) used in the estimate. 
We also draw horizontal lines at the predicted value.}{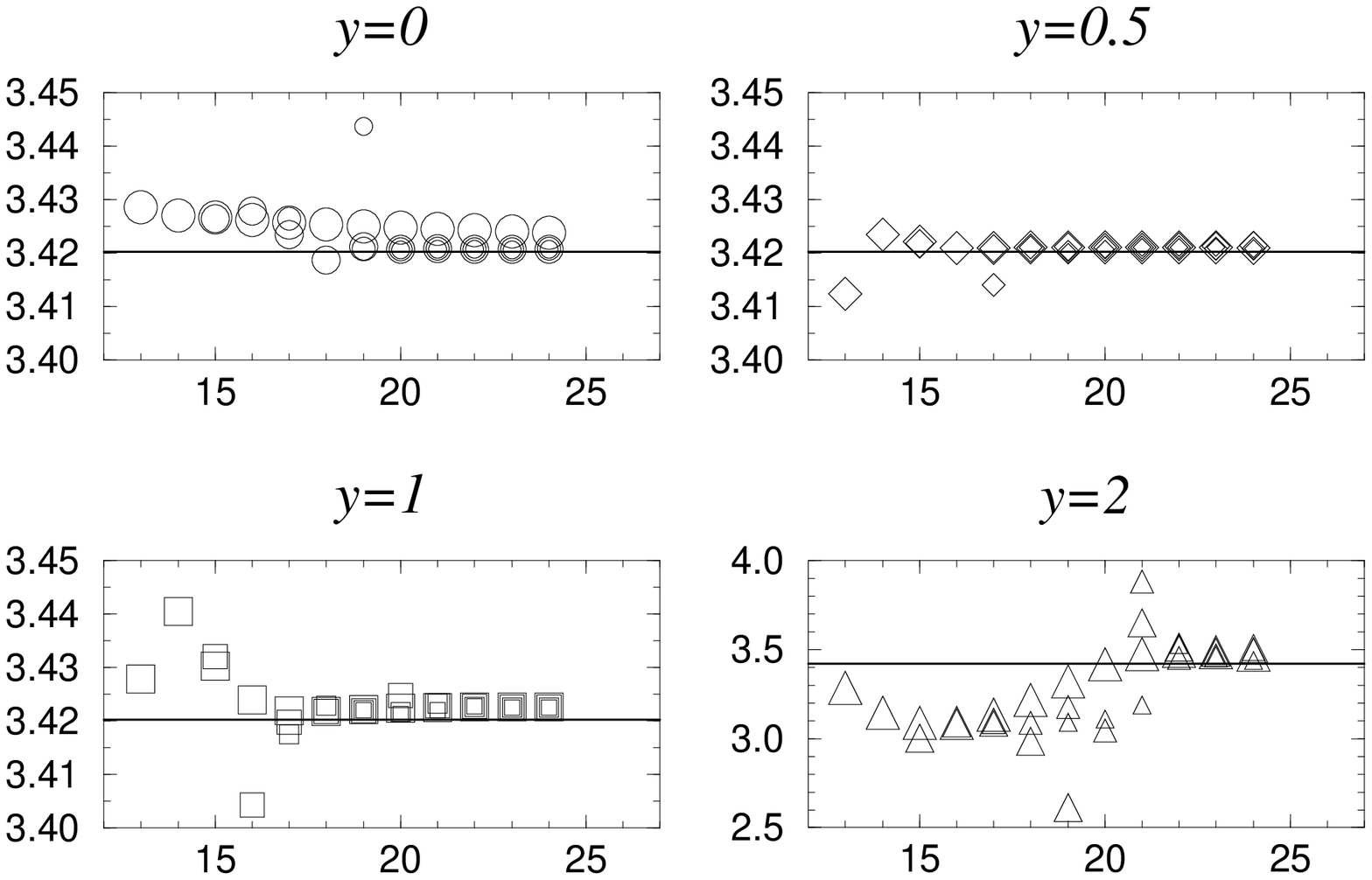}{12.truecm} 
\figlabel\restang

Our results are represented in Fig.~\restang\ and the corresponding
quantitative values listed in Table 7 below. We find a clear evidence
that the configuration exponent is independent of $y$, indicating that
the universality class of tangent meanders and that of meanders are
the same, as predicted in subsection 2.3.  
\medskip
\vbox{
$$\vbox{\font\bidon=cmr10 \bidon
\offinterlineskip
\halign{\quad \hfill # \tv \tv & \hfill # \hfill \tv
& \quad \hfill # \hfill \tv \tv & \quad \hfill # \hfill \tv \tv \cr
$y$ & $\alpha_{\rm theor.}$ & $\alpha_{\rm num.}(y)$ & $R_{\rm num.}(y)$\cr
\noalign{\hrule}
0 & ${29+\sqrt{145}\over 12}=3.4201328\cdots$ & 3.4207 & 3.50184 \cr
1/2 & ${29+\sqrt{145}\over 12}=3.4201328\cdots$ & 3.4208 & 6.188 \cr
1 & ${29+\sqrt{145}\over 12}=3.4201328\cdots$ & 3.422 & 8.735 \cr
2 & ${29+\sqrt{145}\over 12}=3.4201328\cdots$ & 3.4 & 13.63 \cr
}}$$
\nobreak
\noindent{\vbox{\baselineskip=12pt \noindent {\bf Table 7:}
Numerical estimates for the configuration exponent and
the activity $R(q)$ per bridges in the case of
tangent meanders with $q=0$, $x=1$ and $y=0$, $0.5$, $1$ and $2$.  
The error is implicitly on the last digit. 
The corresponding theoretical value is also listed.}}}
\medskip

\newsec{Discussion and conclusion}

In this paper, we have presented theoretical and numerical evidence for a
number of  exact meandric configuration exponents.
Although the sequence of physical arguments leading to the exact predictions
is by no means a mathematical proof, the numerical evidence is compelling.
Let us add a few comments on possible generalizations of our work. 

\subsec{Ranges of validity and transitions}

\fig{The range of validity of the exact prediction
for the meander configuration exponent in the general
case of a weight $n_1$ per river
and $n_2$ per road. In addition to the criticality constraint
that $0\leq n_i\leq 2$, we have represented the $c(n_1,n_2)=1$ curve
(joining the points $(n_1=2,n_2=1)$ and $(n_1=1,n_2=2)$, and passing by
$(n_1=\sqrt{2},n_2=\sqrt{2})$),
beyond which the meanders are in a branched polymer phase. We have also
represented by a dashed line the location of the winding transition
joining the point $(0,q_{\rm c})$ to the point $(1,2)$.}{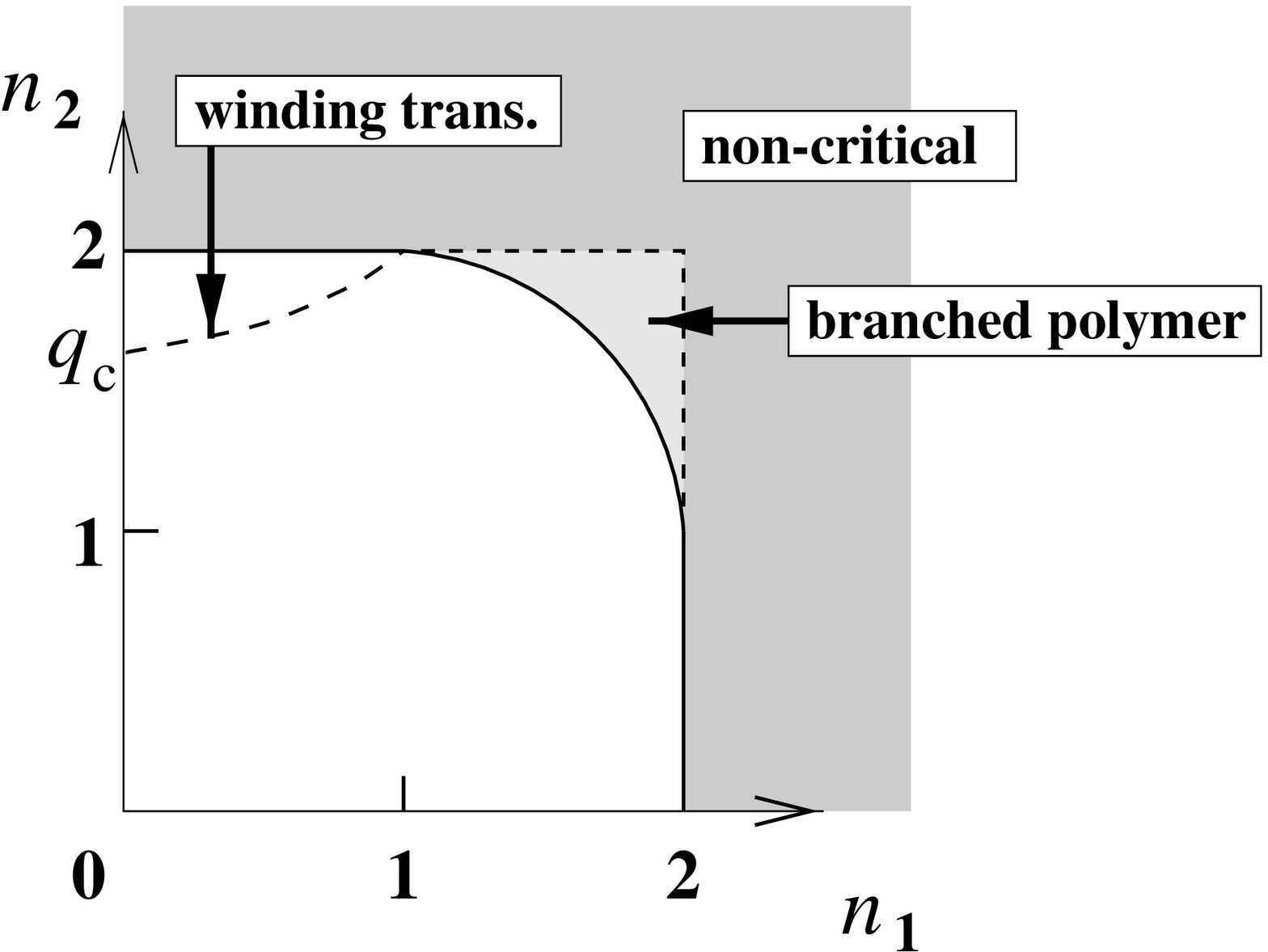}{8.cm}
\figlabel\range

The physical arguments presented above allow to go much farther than
just the case of one or two rivers.
Indeed, we have seen that meanders are a particular case of the
GFPL$^2(n_1,n_2)$ model at $n_1=n_2=0$ with $y=0$.
Let us define multi-river and multi-road meander polynomials
of order $n$, $m_n(n_1,n_2)$, by the
following expansion of the partition function $Z_{\rm GFPL}$ \pfgra\ for $y=0$
\eqn\genmulti{ Z_{\rm GFPL}(n_1,n_2;x)= \sum_{n\geq 1} {x^{2n} \over 4n}
m_n(n_1,n_2) \ .}
Then we can as well predict the following large $n$ asymptotic behavior
\eqn\mepoas{ m_n(n_1,n_2) \sim C(n_1,n_2) {R(n_1,n_2)^{2n} \over n^{\alpha(n_1,n_2)}}\ ,
\qquad \alpha(n_1,n_2)=2-\gamma(c(n_1,n_2)) \ ,}
where $c(n_1,n_2)$ is the central charge of the DPL$^2(n_1,n_2)$ model \newchar,
and $\gamma(c)$ is as in \kpz.
The general prediction \mepoas\ was actually proved in the case
$n_1=1$ and arbitrary $0\leq n_2\leq 2$
in \CK, by solving a particular matrix model.

Note however that the range of validity of \kpz\ imposes that
$c(n_1,n_2)\leq 1$. This gives the total range of validity of Fig.~\range.
Outside of this range, we must consider the two following cases:

\item{(i)} One of the $n_i>2$: then the DPL$^2$ model is no longer critical, and we have no
field theoretical prediction as to the value of the configuration exponent. It has been
shown however, using matrix model techniques, that in the resembling case of
the so-called O$(n)$ model, the critical exponent takes a constant value $\alpha=3/2$
independently of $n>2$ \EKR.

\item{(ii)} Both $n_1,n_2\leq2$ but $c(n_1,n_2)>1$: one encounters the well-known
``c=1 barrier" phenomenon in two-dimensional quantum gravity \BAR.
The corresponding theories
are dominated by configurations of surfaces with long fingers (branched polymer phase
of quantum gravity). Remarkably, it was shown that $\gamma=1/2$ throughout this
phase, leading also to a constant exponent $\alpha=3/2$.

\noindent
Hence we expect (for different reasons) that $\alpha=3/2$ identically outside of the
range of Fig.~\range, for $n_1,n_2\geq 0$.
On the ordinate axis of Fig.~\range, where $n_1=0$, we recover for meanders the 
announced transition at 
$n_2=q=2$, beyond
which the DPL$^2$ model is no longer critical; numerical results however get poorer
as we approach this point and we could not confidently analyze it numerically
so far. 
In the case of semi-meanders, 
we have predicted an earlier ``winding" transition point at 
$n_2=q_{\rm c}=2\cos\pi(\sqrt{97}-1)/48)$. More generally,
the configuration exponent for multi-component
and multi-river semi-meanders (one half-line plus an arbitrary number of river loops) 
is predicted to be
\eqn\predism{ \eqalign{
{\bar \alpha}(n_1,n_2)&= \alpha(n_1,n_2)-1+2 \Delta_1(n_1,n_2) \cr
&=1+{1\over 24}\big(\sqrt{25-c(n_1,n_2)}+\sqrt{1-c(n_1,n_2)}
\big) \sqrt{6(1-e_1)-4c(n_2)}\ , \cr    }}
where $c(n_2)=1-6e_2^2/(1-e_2)$ and $n_i=2\cos \pi e_i$. 
The range of validity of \predism\ is smaller than
the domain of Fig.~\range, as it is delimited by the curve $3(1-e_1)/2+6 e_2^2/(1-e_2)=1$,
with $n_i=2\cos \pi e_i$ (it is represented in dashed line on Fig.~\range). 
The latter corresponds to a winding transition as explained
above, where the number of pieces of road winding around the source of the semi-infinite
river becomes relevant. Above that curve, we expect the semi-meander configuration exponent
to identically vanish.

\subsec{ Extensions}

Extending the generalization of the previous subsection, we could consider {\it both}
complicated river geometries and keeping $n_1,n_2$ finite. This requires some care
when dealing with the electromagnetic operators creating river vertices, as several
geometries might correspond to the same correlators. Also extra care should be
exercised when imposing that the meandric objects be {\it connected}: in the
multi-river and multi-road case, a meander must be globally connected, but
can have disconnected rivers or roads. We may also define higher-genus meandric
numbers~\DGG\ which can also be enumerated by transfer matrix techniques, 
for instance on a torus.

As discussed before, we may also consider the coupling of the FPL$^2$ model to 
Eulerian gravity, namely by summing over genus zero tetravalent graphs with
only faces of even valency. Then all the above predictions are expected to
still hold, except that we must take the formula \centwo\ for the central charge,
that remains one unit above that of the DPL$^2$ model as we have ensured the 
vertex-bicolorability of the graphs we sum over \EKN. 
Note that, in this Eulerian case, the b-vertex of Fig.~\twovert\ is now
relevant and we {\it must} take $y\neq 0$ to get the above shifted central
charge. 
The range of validity of Fig.~\range\ is reduced in this case to the zone 
of the square $[0,2]\times[0,2]$ delimited by the curve $c=1$ joining the point
$(2,q^\star)$ to the point $(q^\star,2)$, with 
$q^\star=2\cos(\pi(\sqrt{13}-1)/6)=0.410135\cdots$. 
One should also be able to check numerically the predictions 
for Eulerian gravity,
using a suitable modification of our transfer matrix technique,
in order to incorporate the vertex-bicolorability.

Another direction of generalization would be to consider more loop colors. For
instance, one can define folding problems of two-dimensional lattices
onto a $d$-dimensional target space, that in turn correspond to fully-packed 
loop models with a given number of loop colors, related to $d$. 
Attaching a different fugacity for loops of different colors, we may generate
different universality classes, described by different conformal theories.
Upon coupling to two-dimensional quantum gravity, these models would correspond
to multi-colored meanders of some kind. 

A final direction of generalization would be to add ``matter" to the meandric
graphs. Indeed, imagine we would like to consider a more involved model for
a compactly folded polymer (protein), by attaching to its nodes a spin variable with
inter and intra-chain Ising-like interactions. Then a simple way of describing 
it would be by first defining this matter spin model on the configurations of the
square lattice fully-packed loop model, and then switching on gravity. If the
matter model is still conformal, this should immediately lead to new
configuration exponents for meanders in the presence of matter.

\subsec{Algebraic exponents}

It is both interesting and sad to notice that the only meandric numbers that
can be calculated exactly with reasonably simple combinatorial formulas are
all in one way or another related to Catalan numbers, and the corresponding 
exponents are always integers or half-integers (take for instance
the case $n_1=0$, $n_2=1$,  then $c=-2$, $\alpha=3$, $\bar \alpha=3/2$).

For general rational values of the $e_i$, we predict however that the exponents are
algebraic numbers, roots of some quadratic equations with integer 
coefficients.  
Remarkably, such exponents are not commonplace in physics, but have emerged in 
some recent works on random walks~[\xref\LOW,\xref\FLM]. 
If we could devise some relation 
between the meander problems and this other type of problems, we would probably
be able to get a mathematically rigorous proof of our predictions.
\medskip
\noindent{\bf Acknowledgements:} We thank O. Golinelli for useful discussions.

\medskip

\listrefs
\end